\documentclass[12pt]{article}

\newcommand{\bfm}[1]{\mbox{\boldmath{$#1$}}}
\newcommand{\beq}{\begin{eqnarray}}
\newcommand{\eeq}{\end{eqnarray}}
\newcommand{\beqs}{\begin{eqnarray*}}
\newcommand{\eeqs}{\end{eqnarray*}}

\usepackage{graphics,epsfig}
\usepackage{latexsym}
\usepackage{amsthm}
\usepackage{amssymb,amsfonts,amsmath}

\begin{document}

\title{Landslides and Mass Shedding on Spinning Spheroidal Asteroids}
\author{D.J. Scheeres\\Department of Aerospace Engineering Sciences\\The University of Colorado Boulder\\scheeres@colorado.edu}
\date{\today}
\maketitle

\begin{abstract}
Conditions for regolith landslides to occur on spinning, gravitating spheroidal asteroids and their aftermath are studied. These conditions are developed by application of classical granular mechanics stability analysis to the asteroid environment. As part of our study we determine how slopes evolve across the surface of these bodies as a function of spin rate, the dynamical fate of material that exceeds the angle of repose, and an analysis of how the shape of the body may be modified based on these results. We find specific characteristics for body surfaces and shapes when spun near the surface disruption limit and develop what their observable implications are. The small, oblate and rapidly spinning asteroids such as 1999 KW4 Alpha and 2008 EV5 exhibit some of these observable traits. The detailed mechanisms outlined here can also provide insight and constraints on the recently observed active asteroids such as P/2013 P5, and the creation of asteroidal meteor streams. 
\end{abstract}

\section{Introduction}

The surface and interior geophysics of small rubble pile asteroids are not fully understood, yet are thought to play an important role in controlling the evolution of these bodies as a function of their spin rates and interactions with other solar system bodies. This study focuses on one specific class of small asteroids, the spheroidal, rapidly spinning asteroids with an equatorial bulge. These bodies are strongly correlated to being a primary in binary asteroid systems, although they also appear frequently as solitary asteroids. The clearest example of this morphology is 1999 KW4 Alpha \cite{KW4_ostro, KW4_scheeres}, which has a pronounced equatorial bulge and a rapid spin rate, with the net gravitational and centripetal acceleration at its equator being near zero. Other well-known examples include Bennu (101955) \cite{nolan_bennu}, the target of the OSIRIS-REx mission, and 2008 EV5 \cite{busch_EV5}, the target of the formerly proposed MarcoPoloR mission. 

In this paper the conditions for landslide failure of regolith on the surfaces of such asteroids are studied. In addition, the associated change in shape of such bodies and the fate of the disturbed regolith are evaluated. These predictions are compared with some known spheroidal-class asteroids to gain insight into the geophysics of such bodies. 
Also investigated are connections between surface slope failures on rapidly spinning bodies and the recent observations of ``active asteroids'', with multiple apparent shedding events occurring over a relatively brief time span \cite{jewitt_active, jewitt_P2010_A, jewitt_P2013_A, hainaut_2013P5}. In this work we find, from a theoretical perspective, that these occurrences of mass shedding could be linked to the morphology of these spheroidal asteroids. 

Our analysis strives for simplicity and thus mainly focuses on a minimal model that can appropriately represent the mechanics that occur for these bodies. Thus, for the shape we will primarily use a sphere. For the regolith properties, we will treat them as cohesionless grains with a specified friction angle that mantle a rigid sphere. For displaced grains undergoing plastic flow, we will make simple assumptions regarding how they will rearrange themselves, specifically assuming that they will preferentially arrange themselves into a flat distribution with zero slope. For the geopotential, we will assume that it can be modeled as a constant density sphere even after deformation. Finally, for grains that are released into orbit, we assume that unless specifically trapped by geopotential curves, that they will be subject to escape. The limitations of these different assumptions will be addressed in our discussions, although our theory will be constructed under their support. 
Finally, despite our simplified analysis we will also show some explicit computations for real asteroid shapes. These will help outline the limitations of our simplifications, and also show how our theory can be modified and extended to more realistic bodies. 

There have been several hypotheses for how loose material can flow across the surface of an asteroid, with a particular emphasis on how equatorial bulges such as seen on binary asteroid primaries such as 1999 KW4 and on solitary spheroidal asteroids such as 2008 EV5 could have formed. In Guibout and Scheeres \cite{guibout_CMDA} it was shown that as an ellipsoidal body spins more rapidly that the ultimate stable location for loose material to settle will always be at the equatorial region. Significantly, this occurs prior to the point where centripetal acceleration exceeds the gravitational acceleration, meaning that loose material is still bound to the surface. In terms of an ellipsoidal shape, the stability transitions between where loose material will settle depends on how the specific shape lies with respect to Jacobi and Maclaurin ellipsoids. 

In Harris et al. \cite{harris_tide}, they study a model with many similarities to the approach we take here. Some similar intermediate results and conclusions are drawn (and are identified later in the text), but ultimately that paper is more focused on the stability and form of static configurations. That paper does identify several hypotheses on how evolutionary spin-up effects can lead to certain configurations, some of which we review and analyze in more depth in the current paper. 
Minton \cite{minton} studied the global shapes of bodies using a more advanced approach for the computation of self-consistent geopotentials for a fixed surface angle of repose. Both of these papers define tools that could be used in conjunction with the current study in future work to develop a consistent model for the deformation and flow of an asteroid surface. 

Other papers have been more focused on the interpretation of existing shape models. In Scheeres et al.\ \cite{KW4_scheeres}, in the context of the shape of 1999 KW4 Alpha, several hypotheses were made on how that object could have gained its bulge. In addition to hypothesizing the surface flow of material to the equator, a hypothesis was made that the bulge could be formed from the in-fall of material that was initially placed in orbit about the body, and potentially driven back to the surface by the presence of the binary secondary body. 
Harris et al.\ \cite{harris_tide} made quantitative direct comparisons between the shape of 1999 KW4 Alpha and their computed radius profiles, pointing out the existence of mid-latitudes at nearly constant slope. 
More recently, Richardson and Bowling \cite{jrichardson_regolith} have studied the relaxation of asteroid surfaces covered with regolith, deriving a model that correlates body density with the observed slope profiles across a body's surface. The focus of that paper is to explain slopes in terms of erosional properties and their migration towards subdued static configurations. In contrast, a key aspect of the current study is to study systems that are on the brink of, or beyond, stability and are in the process of ``falling apart.'' 

There have also been investigations into related phenomenon that take a direct approach to the modeling of a body as a rubble pile. In Walsh et al. \cite{walsh_nature, walsh2012} they model a proto-binary body as a collection of equal sized boulders and simulate its response as it is spun to high spin rates. In their simulations they saw the transport of boulders from the pole down to the equator where they would be flung off into orbit and contribute to the creation of a secondary. Using a different modeling technique S\'anchez and Scheeres \cite{sanchez_icarus} also explored the effect of friction and initial shape on the manner in which a rubble pile asteroid will deform and shed material. For both of these approaches, a limiting factor is that the rubble pile body components are essentially decameter sized boulders, and thus do not provide a high-resolution simulation of how centimeter sized and smaller grains would flow across the surface of an equivalently sized-asteroid. Jacobson and Scheeres \cite{jacobson_icarus}, using a simple model for asteroid fission and evolution, studied how the splitting of components and their subsequent evolution could lead to systems with a fast-spinning primary. 
A different line of investigations have been pursued in \cite{fahnestock_tide, harris_tide}, modeling the surface motion of particles on a spheroidal binary primary as perturbed by its secondary member. These studies have been more focused on how these interactions can cause a binary system to expand through the transfer of angular momentum. While relevant for the evolution of these bodies, the issues that are dealt with herein are more focused on the behavior of bulk materials, and not the system-wide response due to limited motion of surface grains. 


There have also been studies that use continuum mechanics models to provide a global characterization of bodies with geophysical parameters appropriate for modeling regolith. 
Holsapple \cite{holsapple_original,holsapple_yorp} has studied the stability limits for self-gravitating, cohesionless ellipsoids characterized by a friction angle, using a Mohr-Coloumb failure criterion. Sharma has also studied the stability of such assemblages using tools from continuum mechanics \cite{sharma_structure}. Relevant for the current study, Hirabayashi \cite{hirabayashi_bilayer} modeled a rigid sphere mantled by a cohesionless regolith, showing that the existence of a more solid core postponed failure of the body. Comparison of results from Holsapple's study and our current work will show that for a body consisting entirely of cohesionless regolith, that global failure occurs at the same spin rate where surface slope failure occurs. The equivalence between these two failure theories is interesting, and drives an important assumption in our current model, which is that the asteroid is comprised of a rigid sphere mantled by cohesionless regolith up to a given depth. This is modeled in \cite{hirabayashi_bilayer}, where it is shown that the global failure limit of such a body will be at a faster spin rate than for a body which has a uniform distribution of regolith throughout.   

Our current study is not fully distinct from any of these other studies, nor need it operate in isolation of these other effects. In many respects, the current work can be seen as an extension of the initial section of the Harris et al.\ paper \cite{harris_tide}. The unique aspect of our current work is that it develops a more detailed and globally consistent prediction of how asteroid surfaces and sub-surfaces may be redistributed and lost from the surface given the basic mechanical forces acting in this peculiar environment. It is crucial to note that future missions and observations of these spheroidal asteroids will help resolve and clarify our understanding of how these bodies evolve. 

The outline of this paper is as follows. We first review the relevant forces acting on a grain of regolith on, under and above the surface of a spinning body. This includes comparisons between a sphere and an oblate ellipsoid. Following this we introduce the geopotential of a body, with a definition that extends from the interior to the exterior region. Several useful concepts that can be derived from the geopotential are then introduced, such as ``sea level'' on a spinning body, altitude as a function of local slope and body shape, orbit equilibria above a spinning body, and the Roche Lobe. Also, several useful results on the volume beneath different radius and altitude profiles are derived. Then we discuss the granular mechanics of a regolith covered spinning sphere, identifying specific transition points where we would expect material to flow and redistribute itself across the surface. Next we discuss the possible orbital fate of such displaced material, in connection with the Roche Lobe on a spinning spheroid. Finally, we make applications of our results to realistic situations and asteroid models, distinguishing between different possible modes of asteroid failure, and discuss possible observable features that could be identified on the surfaces of bodies that have undergone such landslides as we describe. Connections between these events and active asteroids are also proposed and outlined. 

\section{The Interior, Surface and Exterior Environment on Spinning Spheroids}

First define the environment above, on and below the surface of a sphere of radius $R$ and surface gravity $g_0$, spinning about a fixed axis with an angular rate $\omega$. Assuming rotational symmetry, define an equatorial axis $\hat{\bfm{x}}$ and a polar axis $\hat{\bfm{z}}$, and measure the location of a particle about the body by a radius $r$ and latitude $\delta$, measured from the equator (see Fig.\ \ref{fig:model}). Then, at any point around the body the radial unit vector is $\hat{\bfm{r}} = \cos\delta \hat{\bfm{x}} + \sin\delta\hat{\bfm{z}}$ and the tangent unit vector, pointing in the positive $\delta$ direction, is $\hat{\bfm{t}} = -\sin\delta \hat{\bfm{x}} + \cos\delta\hat{\bfm{z}}$. A grain located a radius $r$ from the center of the body is then inside the body if $r < R$, on the surface of the body if $r=R$, and is above the body surface if $r>R$. 

Technically, we assume that the asteroid is a two-component model, consisting of an inner, rigid sphere of radius $R-H_{reg}$ mantled by cohesionless regolith with a depth $H_{reg}$. This assumption allows us to only consider the failure of the surface regolith, and not be concerned with the global failure of the body. This may be a strong assumption, but it is one which can be probed in the future. It will be seen that the ultimate change in radius of the body is relatively modest, and thus that for the current work this assumption is reasonable. 

Several normalizations are introduced to be used throughout the paper: a mass scale equal to the mass of an individual grain, $m$, a length scale equal to the body radius, $R$, and a time scale equal to the inverse of the angular rate $n = \sqrt{ g_0 / R }$. The parameter $n$ is equal to the angular rate that an object has when in orbit at the surface of the sphere. Introduction of these normalizations yields the result that $r\le 1$ for points within and on the body and $r > 1$ for points in the exterior. The angular rate is also normalized as $\tilde\omega = \omega / n$, and is less than 1 for body spin rates below the point where material will be lofted from the surface of the spinning body, which we call the surface disruption rate. Finally, the acceleration scale is then $R n^2 = g_0$, the gravity at the surface of the sphere.

\subsection{Surface Forces}

A stationary regolith particle on or beneath the surface of a sphere will be subject to the force of gravity from all of the mass below its radius, equal to $-m g_0\left(\frac{r}{R}\right)\hat{\bfm{r}}$, a centrifugal force $m \omega^2 r\left[ \cos^2\delta \hat{\bfm{r}} - \cos\delta\sin\delta \hat{\bfm{t}}\right]$, the overbearing pressure from regolith above its location, and normal and frictional forces of the surrounding media acting on the grain. If there are frictional forces between the grain and its surroundings these can hold the grain in place, balancing all the forces and accelerations. This paper will study conditions under which these assumptions no longer hold and plastic deformation and transport of material can occur. 

\begin{figure}[h!]
\centering
\includegraphics[scale=0.35]{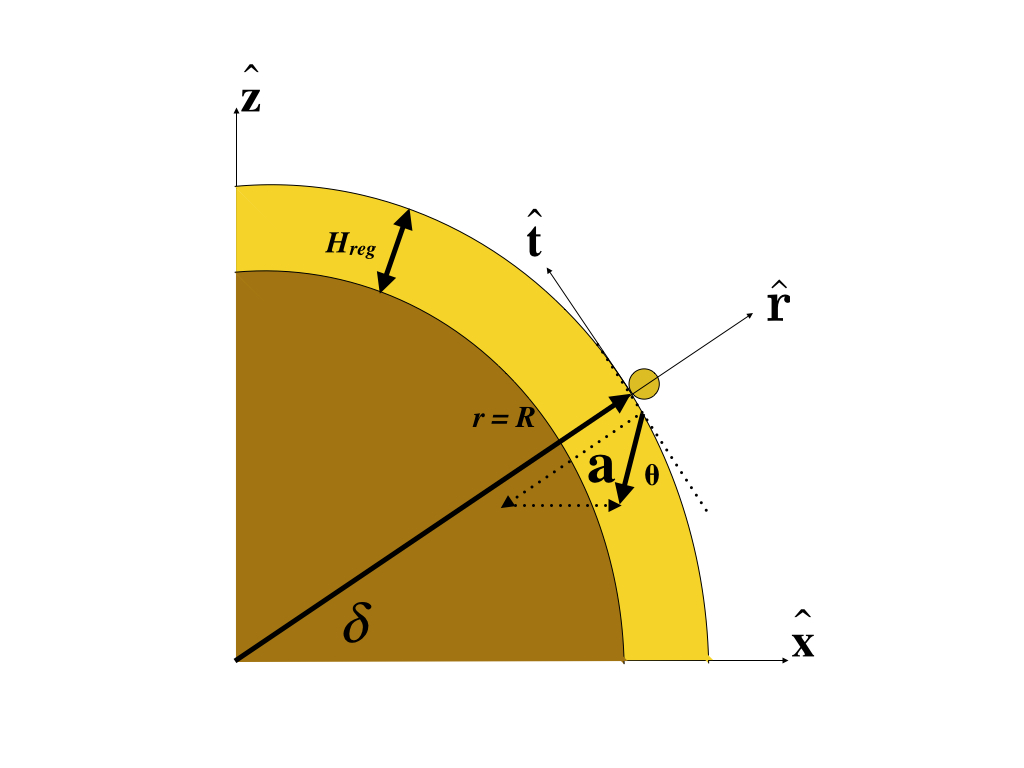}
\caption{Cartoon showing the surface geometry, accelerations and slope.}
\label{fig:model}
\end{figure}

Combining these forces, dividing by the regolith grain mass, and scaling with $g_0$, defines the normalized radial and tangential acceleration at any point in the body, respectively, 
\beq
\bfm{a}_r & = & - \left[ 1 -  \omega^2 \cos^2\delta \right] r \hat{\bfm{r}} \\
\bfm{a}_t & = & -\tilde\omega^2 r \cos\delta \sin\delta \hat{\bfm{t}}
\eeq
In a static situation these forces are counteracted by the frictional and normal forces mentioned previously. 
A few immediate comments can be made. At a given radius the radial acceleration will in general be less than the gravitational acceleration, with the exception occurring at the pole of the body. The tangential acceleration always acts away from the pole and towards the equator. This can become modified for an oblate ellipsoid (discussed later), however this trend generally holds for large enough spin rates. The total acceleration that a regolith grain will experience, and that must be countered by the surrounding media to keep the grain in place, is then
\beq
a & = & r \sqrt{ 1 - (2-\tilde\omega^2) \cos^2\delta \tilde\omega^2 }
\eeq
Thus, if $\tilde\omega = 0$, the body is not spinning and the acceleration is constant across the surface, and if $\tilde\omega = 1$ then loose material at the equator (at all levels of radius) will have no net acceleration acting on them, although regions at non-equatorial latitudes will still have net attractive accelerations acting on them. Figure \ref{fig:model} shows a cartoon of the surface coordinate, accelerations and net slope. 

\subsection{Surface Slopes}

A surface can be defined, in part, by its local slope, which measures the angle of the topography to the total acceleration acting at the surface. The tangent of the slope angle $\theta$ equals the ratio of the tangential acceleration over the negative radial acceleration. 
\beq
	\tan\theta & = & \frac{\tilde\omega^2 \cos\delta \sin\delta}{1 - \tilde\omega^2\cos^2\delta} \label{eq:slope} \\
	& = & \frac{\tilde\omega^2 \tan\delta}{1 - \tilde\omega^2 + \tan^2\delta} \label{eq:slope2} 
\eeq
and is independent of radius. 
The slope is always zero at the equator and the pole (for $\tilde\omega < 1$) and is always positive leading away from the equator. This means that the equator of a spinning sphere is the lowest point on the surface and that loose material will preferentially flow towards this point. 

The maximum slope on the surface is found by computing $\partial\tan\theta / \partial \delta = 0$ and solving for the latitude $\delta^*$ where this is satisfied. Carrying out this computation yields the maximum value of slope as a function of $\tilde\omega$
\beq
	\tan\theta^* & = & \frac{\tilde\omega^2}{2\sqrt{1-\tilde\omega^2}}
\eeq
which occurs at a latitude defined by $\tan\delta^* = \sqrt{1-\tilde\omega^2}$. Thus the latitude of maximum slope occurs at $45^\circ$ latitude for nearly zero spins and moves down towards the equator for more rapid spins. At the same time, the maximum value of the slope increases with increasing spin rate, until it reaches a value of $90^\circ$ at the equator for $\tilde\omega = 1$. The locus of maximum spin rates as function of latitude can be found by substituting $\omega = \sqrt{1-\tan^2\delta^*}$ into the equation for $\tan\theta^*$. Doing so yields the simple relation for the locus of maximum slopes as a function of latitude
\beq
	\theta^* & = & \pi/2 - 2 \delta^*
\eeq

\begin{figure}[h!]
\centering
\includegraphics[scale=0.35]{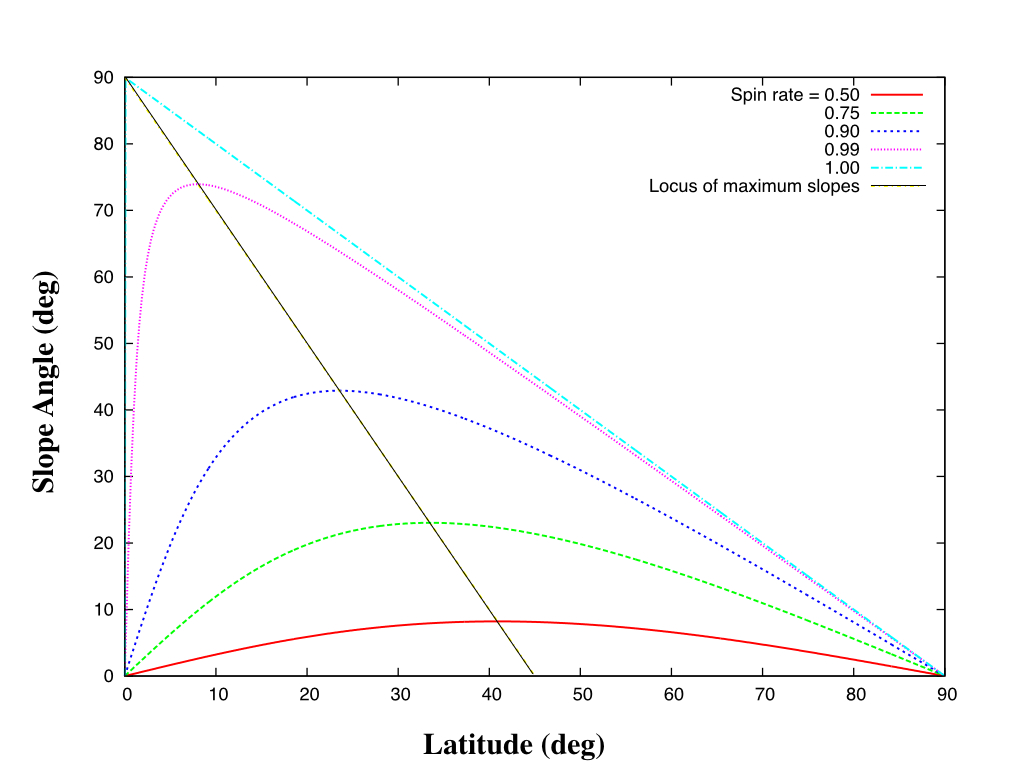}
\caption{Surface slopes across a sphere as a function of spin rate. }
\label{fig:slopes}
\end{figure}

At extreme spin rates it is also instructive to note that the slope takes on a specific structure. Let $\tilde\omega\rightarrow 1$ to find $\tan\theta = \cot\delta$, or $\theta = \pi/2 - \delta$. Thus, even though the slope goes to $90^\circ$ at the equator, it is always less than the colatitude across the surface of the body. Specifically, in the region of the pole the slope will always be low and close to zero, independent of the spin rate of the body. Figure \ref{fig:slopes} shows the surface slopes on a spinning sphere as a function of latitude for a range of different $\tilde\omega$ values. We note that it is essentially the same as the Fig.\ 1 in Harris et al. \cite{harris_tide}, yet we keep it here for completeness. 

\subsection{Comparisons with Oblate Spheroids}

It is instructive to compare this result with a similar calculation for ellipsoids to ascertain the limitations that are introduced by using a sphere as the basic shape. It is well known that slopes over an ellipsoid will have distinct differences as compared to a sphere. In \cite{guibout_CMDA} the stable resting points of loose material on the surface of an ellipsoid is studied and shown to be controlled by where a given ellipsoid lies relative to the Jacobi and Maclauren sequence of ellipsoids in terms of shape and spin rate. One conclusion from that paper is that for a fast enough spin rate (but less than the shedding limit) the geopotential low always lies at the equator. This will prove to be a key feature for the sphere, and thus justifies its use for the current study. 

\begin{figure}[h!]
\centering
\includegraphics[scale=0.25]{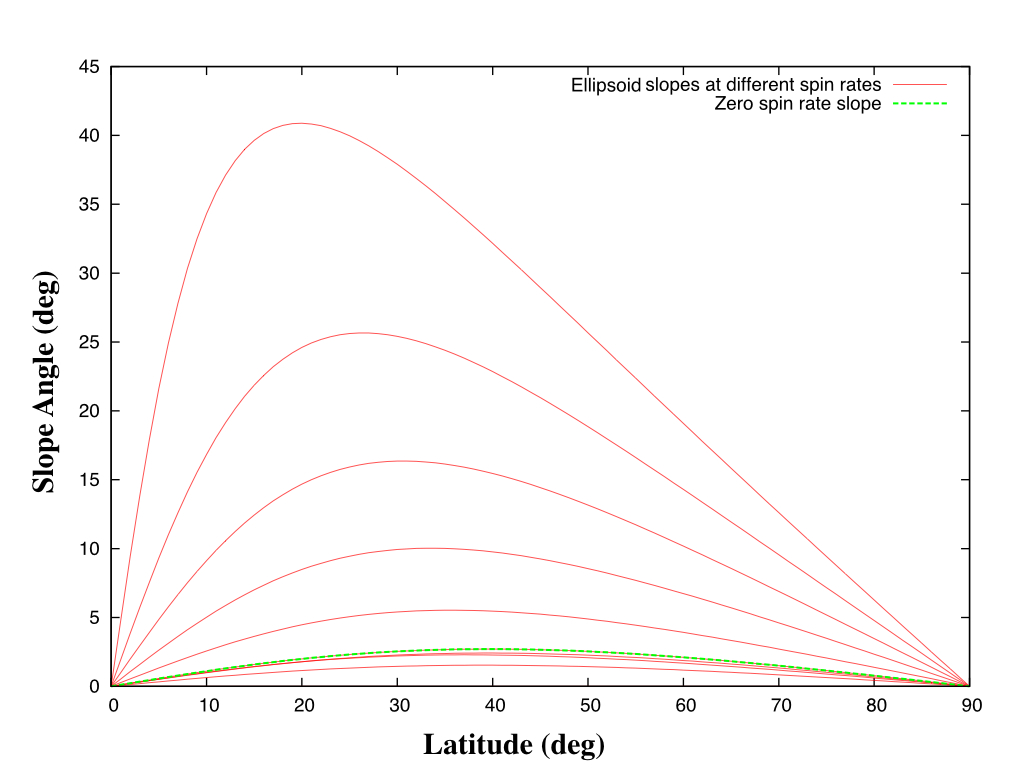}
%
\includegraphics[scale=0.25]{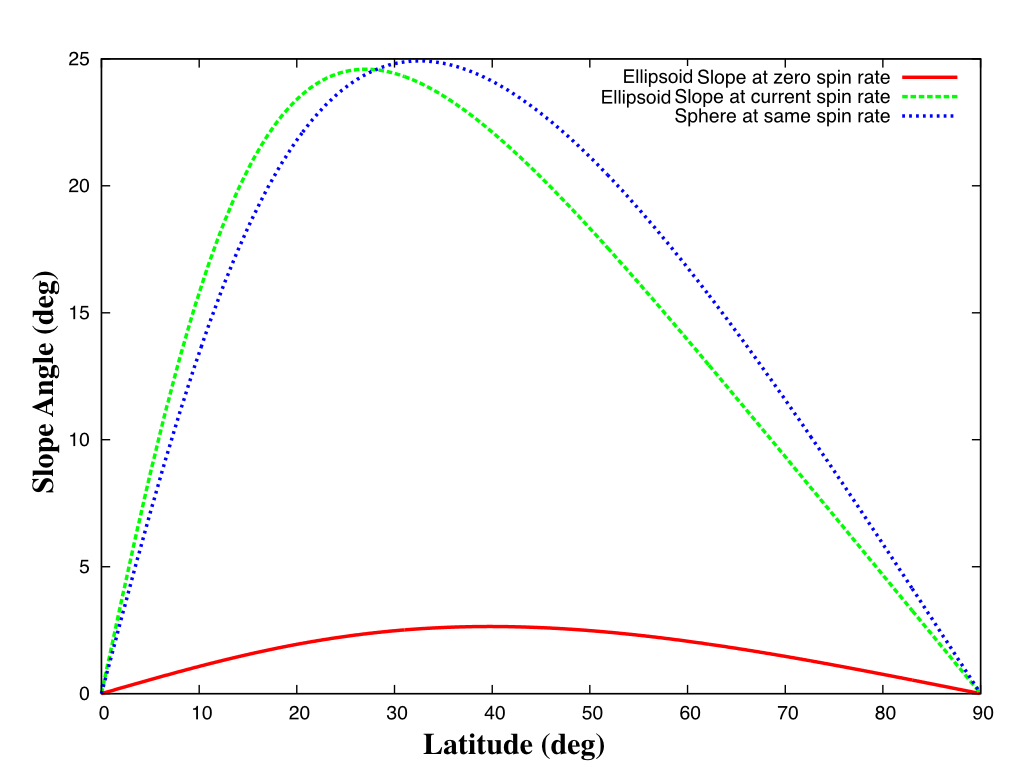}
\caption{Top: Surface slopes over the surface of a nearly oblate ellipsoid.
Bottom: Surface slope comparisons between a nearly oblate ellipsoid and a sphere at a spin rate of $\tilde\omega = 0$ and $ 0.793$.}
\label{fig:bennu2}
\end{figure}

Figure \ref{fig:bennu2} show plots for the surface slope of an ellipsoid with relative dimensions $1\times 0.966 \times 0.889$, chosen as a representative set of values for an ellipsoidal body. The top panel shows surface slopes for the ellipsoid along the maximum and minimum semi-major axes of that body at values of $\tilde\omega$ ranging from 0 to 0.9 in steps of 0.1 (see \cite{scheeres_asteroid_book} for the computational procedure). The zero spin rate has non-zero slopes over the surface, due to the ellipsoidal shape, but as the spin rates are increased the maximum slope first lessens and then starts to increase again. This transition occurs at a relatively modest spin rate of 0.4. After this transition, the surface slopes are qualitatively similar to the same slopes seen on the surface of a sphere. The bottom panel shows a direct comparison of the ellipsoid surface slope and a sphere's surface slope at the same normalized spin rate, noting that they are quite similar. In fact, based on this close similarity for surface slopes at a rapid spin rate, in the following we only consider the surface slopes on a spinning sphere, enabling almost all of the computations to be carried out analytically. 

\clearpage

\subsection{Forces off the Surface}

For points not in or on the surface of the asteroid, the centripetal acceleration formula remains the same while the gravitational acceleration reverts to the classical Keplerian version. 
The normalized radial and transverse accelerations are 
\beq
	\bfm{a}_{r} & = & - \left[ \frac{1}{r^2} - \tilde\omega^2 r \cos^2\delta \right]  \hat{\bfm{r}} \label{eq:arx} \\
		\bfm{a}_{t}  & = & - \tilde\omega^2 r \cos\delta\sin\delta \hat{\bfm{t}} \label{eq:atx} 
\eeq

Of special interest is the point off of the asteroid surface where there is no net acceleration acting on a grain in the rotating frame. This is found by solving for the point $(r,\delta)$ which makes both of the accelerations zero. Considering the second equation, either $\delta = 0,\pi/2$ would work, however at $\delta = \pi/2$ the first equation is always non-zero. Taking $\delta = 0$, Eqn.\ \ref{eq:arx} is solved to get the equilibrium radius that defines the synchronous circular orbit at a given spin rate
\beq
	r^* & = & \frac{1}{\tilde\omega^{2/3}} 
\eeq
Note that $r^* > 1$ for $\tilde\omega < 1$, and that when $\tilde\omega = 1$ the equilibrium point touches the surface of the spinning body at the equator. In \cite{hirabayashi_kleo} this condition is defined as surface disruption, as material can freely lift off the surface when this is satisfied. However, long before this limit is reached we will show that granular surfaces will have failed already. 

Another important concept is to define the distance from the surface of the sphere at which a grain will experience a net outwards acceleration, which will be an important limiting constraint on the extent to which material can be shifted. 
The most conservative condition can be identified as when $a_r \ge 0$, which would require cohesive strength for a grain to remain attached to the surface. Solving this equation yields the limiting radius as a function of latitude
\beq
	r_{r} & = & \frac{1}{(\tilde\omega\cos\delta)^{2/3}}
\eeq
Note that for low latitudes this is approximately $r_r \sim r^*$. 

A stricter constraint can be derived based on when the centripetal acceleration overpowers the horizontal gravity component, as at this point a grain will preferentially migrate outwards and away from the body if disturbed. For this case the acceleration along the $x$-axis is found by combining Eqns.\ \ref{eq:arx} and \ref{eq:atx} 
\beq
	a_x & = & \bfm{a}_r \cdot \hat{\bf x} + \bfm{a}_t \cdot \hat{\bf x} \nonumber \\
		& = & r\omega^2 \cos\delta - \frac{1}{r^2} \cos\delta
\eeq
The limiting condition for $a_x = 0$ is then independent of $\delta$ and is  
\beq
	r_x & = & \frac{1}{\tilde\omega^{2/3}} = r^*
\eeq
This is a constant radius limit that holds for all latitudes, and is applied later when we consider the redistribution of material. 

\section{The Geopotential and Associated Constraints}

The geopotential is a fundamental function for the current analysis. For completeness, and for application to the asteroid environment, it must be defined from the interior to the exterior. In addition, we derive several applications of the geopotential to define important concepts that will be used later in our model. 

\subsection{The Geopotential} 
The normalized geopotential for a particle above, on and under the surface of a simple, spinning spherical body can be described as 
\beq
	V(r,\delta) & = & - \frac{1}{2} \tilde\omega^2 r^2 \cos^2\delta + U(r) \\
	U(r) & = & \left\{ \begin{array}{cc} - \frac{1}{r} & r > 1 \\ \\ - \frac{1}{2} \left[ 3  - r^2 \right] & r \le 1 \end{array} \right.
\eeq
For sub-surface considerations the geopotential is
\beq
	V_{sub}(r,\delta) & = & \frac{1}{2} \left[ \left(1-\tilde\omega^2\cos^2\delta\right) r^2 - 3\right] 
\eeq
and for external considerations the geopotential is 
\beq
	V_{ext}(r,\delta) & = & - \left[ \frac{1}{r} + \frac{1}{2} \tilde\omega^2 r^2 \cos^2\delta \right]   
\eeq
For both cases the geopotential can be used to generate the accelerations acting on a grain as $a_r = - \partial V/\partial r$ and $a_t = - 1/r \ \partial V / \partial \delta$. 

\subsection{Sea-Level Definition and Altitude}

At the surface of the body, $r = 1$, the minimum value of the geopotential occurs at the equator, defining the concept of ``sea-level'' and providing a reference value of the potential as a function of spin rate, 
\beq
	V_{sl} & = & V_{sub}(1,0) \\
	& = & \frac{1}{2} \left[ \left(1-\tilde\omega^2\right) - 3 \right] 
\eeq
This position on the equator remains the lowest point on the surface of a sphere for all spin rates less than the disruption spin rate.
This can be used to define an ``altitude'' relative to sea level, were material above sea level will energetically be able to move down to a lower geopotential value if given sufficient motive force. 
To define the relative altitude, equate $V_{sl} = V_{sub}(r_{sl},\delta)$ and solve for $r_{sl}$ to find
\beq
	r_{sl} & = & \sqrt{\frac{1-\tilde\omega^2}{1-\tilde\omega^2\cos^2\delta}} \label{eq:sealevel}
\eeq
Radii and locations where $r > r_{sl}$ are then above sea-level, and those below this point are below sea-level.  One immediate observation is that all surface material measured away from the equator is above sea-level for any non-zero spin rate, as $r_{sl} < 1$. Specifically, the ``altitude'' of the surface of the spinning sphere is
\beq
	H_{surf} & = & 1 - \sqrt{\frac{1-\tilde\omega^2}{1-\tilde\omega^2\cos^2\delta}} 
\eeq
Meanwhile, the altitude of an arbitrary point within the spinning sphere can be computed as 
\beq
	H & = & r - \sqrt{\frac{1-\tilde\omega^2}{1-\tilde\omega^2\cos^2\delta}}  
\eeq

\begin{figure}[h!]
\centering
\includegraphics[scale=0.35]{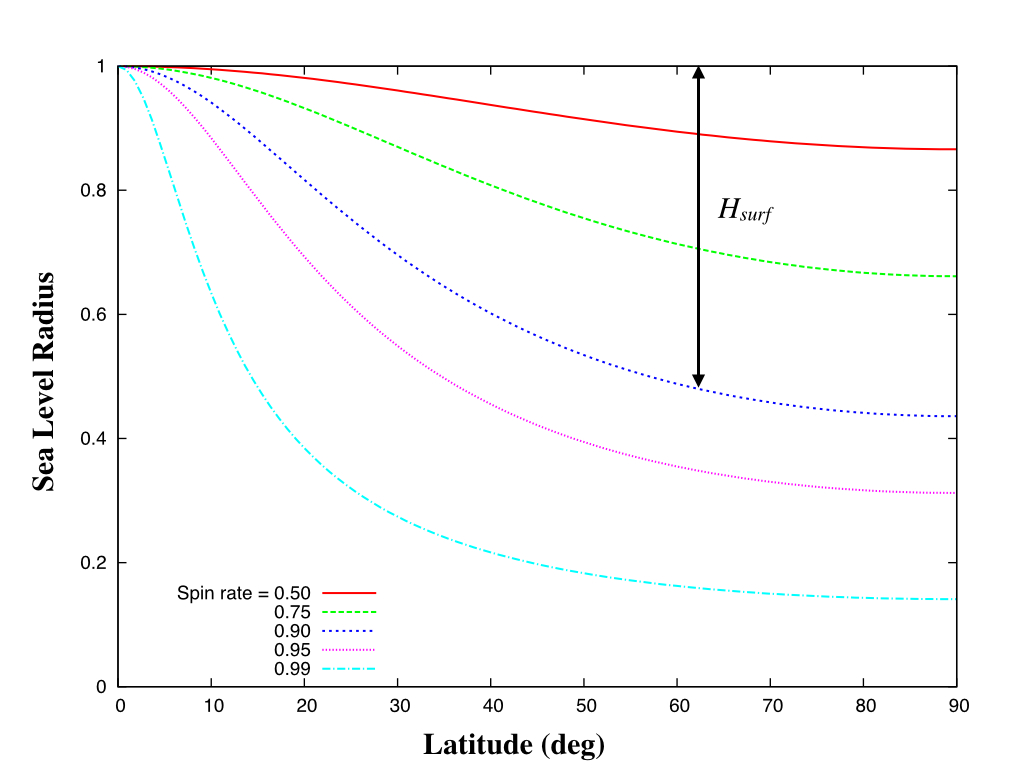}
\caption{Sea-level radius as a function of latitude for different spin rates. The surface altitude of a specific point is indicated. }
\label{fig:sealevel}
\end{figure}

Figure \ref{fig:sealevel} shows the sea-level radius within the body as a function of latitude for different spin rates. We note that the polar region is always at the highest altitude above sea-level. 

\subsection{Altitude as a Function of Slope}

An alternate definition of altitude can be developed using the effective surface slope of the spinning sphere and a physically changing slope on the surface or within the body. 
Here we note that there are two concepts of slope that we wish to combine. The first is the apparent slope at a given point $(r,\delta)$ in the body, $\tan\theta$, and defined by Eqn. \ref{eq:slope}. This slope defines the apparent change in height at a constant radius for an increasing or decreasing latitudinal direction. Added to this variation is the actual variation in radius as a function of latitude, which defines the geometric slope and is equal to $\frac{1}{r} \frac{d r}{d\delta}$. The sum of these two, the apparent and geometric slopes, defines the total slope of an altitude profile. Of particular interest are profiles that yield a constant slope angle, possibly zero. If we define a slope angle of $\phi$, then the equation that relates this to the apparent slope and the geometric slope is
\beq
	\tan\phi & = & \tan\theta + \frac{1}{r} \frac{d r}{d\delta} 
\eeq

This equation can be transformed to a differential equation for the radius 
\beq
	\frac{1}{r} \frac{dr}{d\delta} & = & \tan\phi - \frac{\tilde\omega^2 \sin\delta\cos\delta}{1-\tilde\omega^2\cos^2\delta}
\eeq
If the specified slope angle $\phi$ is held constant, separation of variables applies and this equation can be integrated between a reference radius and latitude, $r_o$ and $\delta_o$, and a specified radius and latitude, $r$ and $\delta$
\beq
	\ln\frac{r}{r_o} & = & (\delta - \delta_o) \tan\phi + \ln \sqrt{\frac{1-\tilde\omega^2\cos^2\delta_o}{1-\tilde\omega^2\cos^2\delta}}
\eeq
which can be rewritten as
\beq
	\frac{r}{r_o} & = & \sqrt{\frac{1-\tilde\omega^2\cos^2\delta_o}{1-\tilde\omega^2\cos^2\delta}} e^{(\delta - \delta_o) \tan\phi} \label{eq:radius_slope}
\eeq
This equation gives the radius profile for a constant slope surface starting from $(r_o,\delta_o)$ as a function of latitude. If we take $r_o = 1$, $\delta_o = 0$ and specify a zero slope condition, $\phi = 0$, the sea-level radius formula defined in Eqn.\ \ref{eq:sealevel} using only the geopotential is recovered. 

\subsection{Roche Lobe and Trapped Surface Material}

The geopotential above the surface can be used to compute the energetics of motion in the vicinity of the asteroid. Here the geopotential is measured relative to its value at the equilibrium point,
\beq
	V_{ext}(r^*,0) & = & - \frac{3}{2} \tilde\omega^{2/3}
\eeq
This value of the geopotential defines the Roche Lobe (c.f., \cite{dobrovolskis, scheeres_asteroid_book}), specifically regions in the space above and even below the surface which are energetically trapped at the asteroid or, conversely, regions which are able to escape from the asteroid surface and reach arbitrary distances away from the body. 

\begin{figure}[h!]
\centering
\includegraphics[scale=0.25]{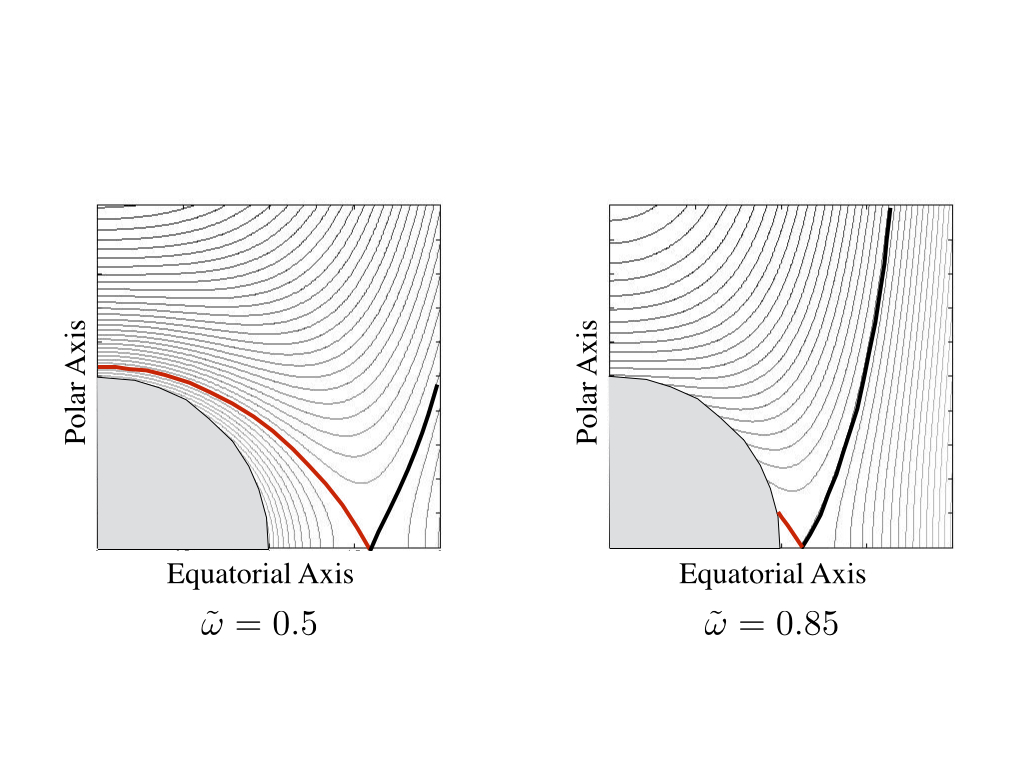}
\caption{Zero-velocity curves for $\tilde\omega = 0.5, 0.85$. Any surface material beneath the red curve will remain trapped even under migration towards the equator. Material above the red curve and on the surface can depart from the vicinity of the asteroid surface.}
\label{fig:zvcurve}
\end{figure}

To map out these regions, consider the modified energy of this system as expressed in the rotating frame. 
\beq
	E & = & T + V(r,\delta)
\eeq
where $T$ is the kinetic energy in the body-fixed frame and is zero when at rest with respect to the asteroid surface. The energy of the equilibrium point can be evaluated as $E^* = -\frac{3}{2}\tilde\omega^{2/3}$. Then regions of allowable motion can be defined using the constraint that $T = E^* - V(r,\delta) \ge 0$ to find
\beq
	\frac{3}{2}\tilde\omega^{2/3} & \le & \frac{1}{2} \tilde\omega^2 r^2 \cos^2\delta + \left\{ \begin{array}{cc}
		\frac{1}{r} & r > 1 \\
		\\
		\frac{1}{2}\left( 3 - r^2 \right) & r \le 1 \end{array} \right.
\eeq
The points which satisfy this inequality and have $r < r^*$ are bound to the asteroid and do not have sufficient energy to depart its vicinity. 
Those points which violate this inequality have sufficient energy to escape through an open region around the equilibrium point, even if they still lie in the region $r<r^*$ (see Fig.\ \ref{fig:zvcurve}). 

The Roche Lobe is defined at the locus of points $(r \le r^*,\delta)$ such that the expression is an equality. For points in the exterior region, the defining Roche Lobe surface of radius as a function of latitude, $\delta$, can be reduced to finding the positive roots of a cubic equation
\beq
	\left(\frac{1}{r_{RL}}\right)^3  - \frac{3}{2}\tilde\omega^{2/3} \left(\frac{1}{r_{RL}}\right)^2 + \frac{1}{2} \tilde\omega^2\cos^2\delta & = & 0
\eeq
which must be found numerically and is only valid when $r_{RL} \ge 1$. For points on and beneath the surface, the condition can be solved explicitly as
\beq
	r_{RL} &  = & \sqrt{ \frac{3(1-\tilde\omega^{2/3})}{1-\tilde\omega^2\cos^2\delta}}
\eeq
and is only valid for $r_{RL} \le 1$. 

Evaluating the geopotential on the surface of the asteroid, with $r = 1$, identifies which regions of the asteroid surface are energetically trapped and which have sufficient energy to leave the vicinity of the asteroid.
\beq
	\cos^2\delta & \ge & \frac{2}{\tilde\omega^2} \left[ \frac{3}{2} \tilde\omega^{2/3} - 1\right] 
\eeq
For $\tilde\omega < (2/3)^{3/2} \sim 0.544\ldots$ the entire surface satisfies the inequality and is energetically bound to the vicinity of the asteroid. Note that this is the usual case for non-spheroidal asteroids (c.f., \cite{hirabayashi_kleo}). Once $\tilde\omega$ is above this limit, the intersection of the asteroid surface with the zero-velocity surface begins to migrate to lower latitudes, initially exposing the polar regions. This means that material at higher polar regions have sufficient energy, if they move from their location, to escape the vicinity of the asteroid. This does not mean that they will, of course, as they may lose or dissipate their kinetic energy as they slide down the surface. Figure \ref{fig:zvcurve} shows the contours of constant geopotential for values of $\tilde\omega = 0.5, 0.85$. At the lower value of $\tilde\omega$ the entire surface is beneath the Roche Lobe, shown in red. For the larger value only a small region of the asteroid surface in the vicinity of the equator is below the energetic limit. As $\tilde\omega\rightarrow 1$ this region shrinks to a point on the surface. 


\begin{figure}[h!]
\centering
\includegraphics[scale=0.25]{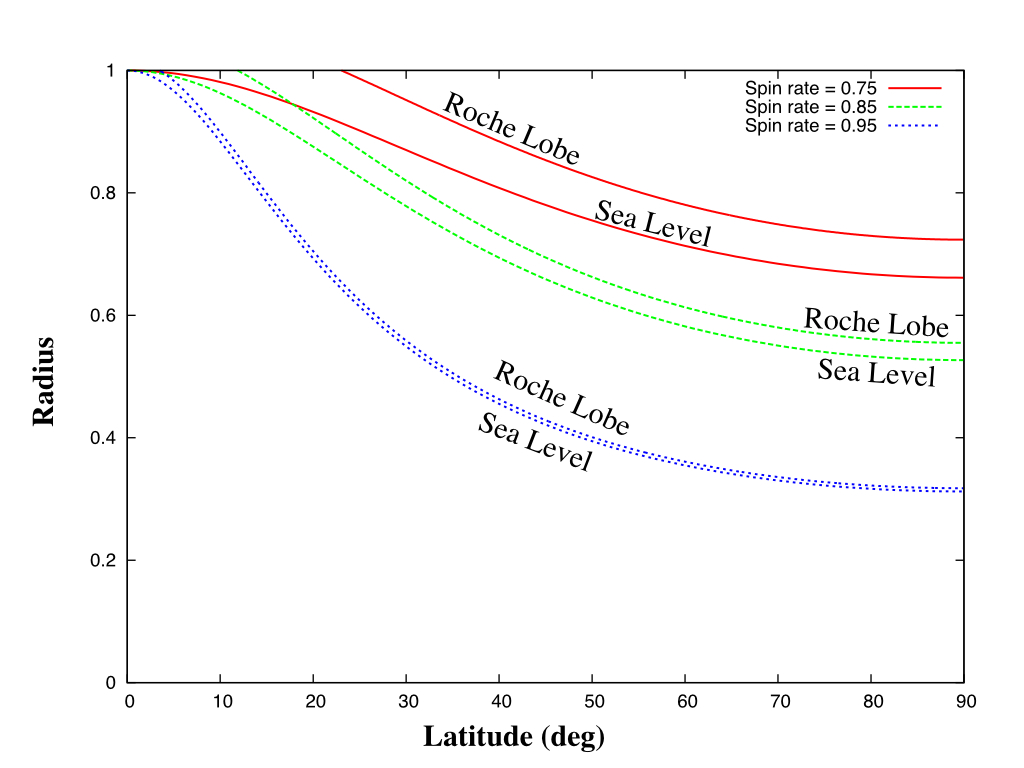}
\caption{Comparison between sea-level and Roche Lobe radii at different spin rates. Sea level always lies beneath the Roche Lobe.}
\label{fig:ZV_SL}
\end{figure}

It is instructive to compare the Roche Lobe that extends within the body to the sea level. Figure \ref{fig:ZV_SL} shows the Roche Lobe and sea level radius profiles at different levels of spin rate. We note that $r_{sl} < r_{RL}$, but that as spin rates increase these radii converge towards each other. 


\subsection{Volume Beneath Radius Profiles}

\begin{figure}[h!]
\centering
\includegraphics[scale=0.3]{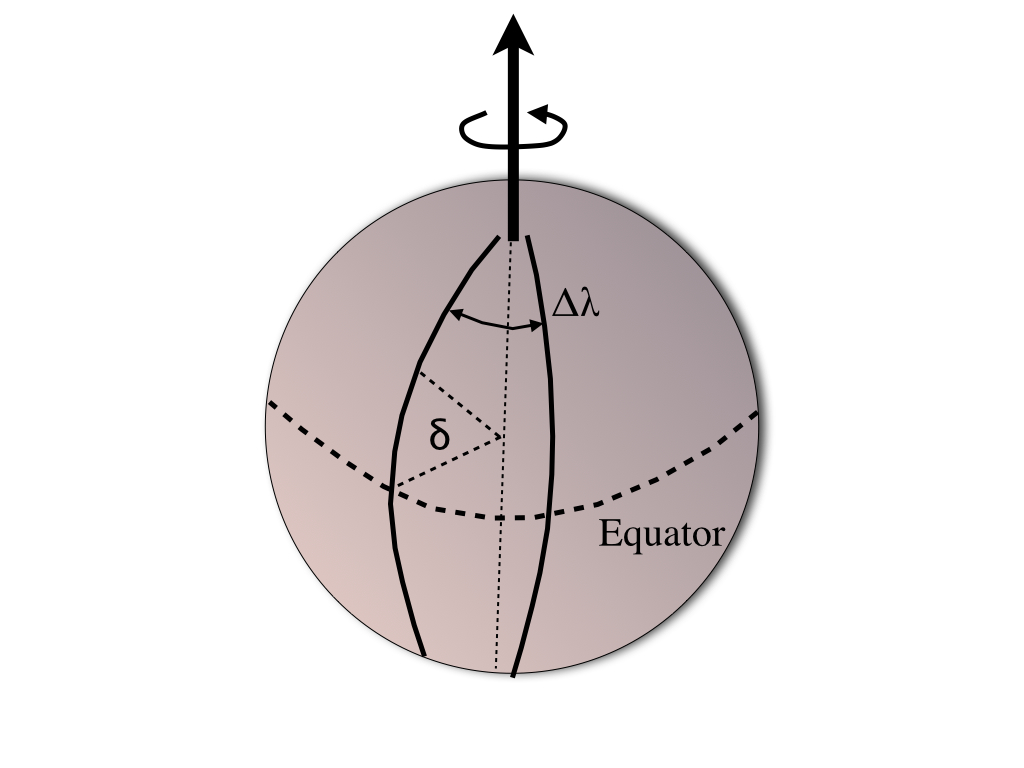}
\caption{Cartoon showing a lune on a spinning sphere.}
\label{fig:lune}
\end{figure}

Finally, the volume below different radius profiles are defined, which is important when considering the redistribution of material. As the model has a symmetry in the longitudinal direction, everything can be formulated in terms of an arbitrary longitudinal width, denoted as $\Delta \lambda$ (see Fig.\ \ref{fig:lune}). Then the volume element at an arbitrary point within the body, for a fixed longitudinal width, is specified as
\beq
	dV & = & r^2 \cos\delta \ d\delta \ dr \ \Delta\lambda
\eeq

The volume is measured from the center of the body to a radius profile defined as a function of latitude. Thus the general volume formula is
\beq
	V & = & \Delta\lambda \ \int_{r=0}^{r(\delta)} \int_{\delta_o}^\delta r^2 \cos\delta \ d\delta \ dr 
\eeq
The radial integration can be immediately performed, yielding 
\beq
	V & = & \frac{1}{3}\Delta\lambda \int_{\delta_o}^\delta r(\delta)^3 \cos\delta \ d\delta \label{eq:volume}
\eeq
Then, depending on the specification of $r(\delta)$, the integral will take on different forms. 

First consider the simplest case, the volume beneath the initial surface of the sphere. Then $r(\delta) = 1$ and the integral has the well-known result
\beq
	V_{surf} & = & \frac{1}{3} \Delta\lambda \left( \sin\delta - \sin\delta_o\right) \label{eq:vsurf}
\eeq
Taking $\delta_o = 0$, $\delta = \pi/2$ and $\Delta\lambda = 2\pi$ yields $V_{surf} = 2\pi/3$, which is the volume of a hemisphere.

The total volume ``beneath'' sea-level is found by substituting the sea-level radius into the integral, and can be evaluated to find
\beq
	V_{sl} & = & \frac{\Delta\lambda}{3} \sqrt{\frac{1-\tilde\omega^2}{1-\tilde\omega^2\cos^2\delta}} \sin(\delta) \nonumber \\
	& = & \frac{\Delta\lambda}{3} r_{sl}(\delta) \ \sin\delta
\eeq
The volume of material above sea-level is then found by subtracting the sea-level volume from the surface volume to find
\beq
	\Delta V_{sl} & = & \frac{\Delta\lambda}{3} \sin\delta \left( 1 - r_{sl}(\delta) \right)  \label{eq:deltasl}
\eeq

The volume beneath a radius profile for a constant, non-zero slope cannot be evaluated analytically. Substituting Eqn.\ \ref{eq:radius_slope} for the radius profile into Eqn.\ \ref{eq:volume} yields 
\beq
	V_{\phi} & = & \frac{r_o^3 \Delta\lambda}{3} \int_{\delta_o}^\delta \left[ \frac{1-\tilde\omega^2\cos^2\delta_o}{1-\tilde\omega^2\cos^2\delta} \right]^{3/2} \cos\delta \ e^{3(\delta-\delta_o)\tan\phi} d\delta \label{eq:vol_const}
\eeq
and must be evaluated numerically. The difference between the total volume to the surface and the volume beneath the constant slope profile is then
\beq
	\Delta V_{\phi} & = & V_{surf} - V_{\phi} \label{eq:deltaphi}
\eeq

\section{Evolution of Loose Regolith}

In this section the above analysis is applied with classical granular mechanics of cohesionless systems to develop predictions and constraints on how regolith on a rapidly spinning spheroid may fail and redistribute itself. The current analysis does not consider the dynamics of the flowing regolith, a topic for future inquiry using detailed granular mechanics simulations. The current analysis is biased towards analytical results and conservative constraints. Nonetheless, these are crucial elements of theory that need to be understood in order to motivate future, more advance modeling and research on the phenomenon of surface failure. 

\subsection{Granular Mechanics Model}

%
In Harris et al.\ \cite{harris_tide} they use a more precise geophysical model for granular materials, the angle of slide and the angle of repose. The former is the maximum slope angle before a material will flow and the latter is the angle at which the material stops its flow. While this model ignores the possible effects of land sliding, where material can continue to flow into a runout region due to inertia, it is a more precise description than we use. Due to difficulties in developing simple and analytic models for volume flow and redistribution we assume the angle of slide and angle of repose are equal, and equals its angle of friction, which is denoted henceforth as $\phi$. When the local surface slopes of a granular pile exceed this value the material will flow, leaving behind a surface of constant slope $\phi$. The theory is relatively simple and can be found in \cite{nedderman}. The beauty of this model is that the flow rules only need to be applied to the free surface of a granular material. Thus, unlike for cohesive grains, there is no need to specify the total weight and forces acting on potential slip planes within the granular pile. Rather, whenever the angle of repose is exceeded at the free surface, the material will flow. 

While the specification of granular mechanics failure is simple, deciding how to model the material that flows towards the equator requires some assumptions. We will assume that mobile regolith will tend to continue to flow down to the lowest point in the geopotential, which will be to the equator. Given the symmetry of our model, such a flow will be met at the equator by an equivalent flow over the other hemisphere. We will make the further strong assumption that the regolith will settle into a zero-slope surface at the equator. In reality, the final surface can take on any local slope at or below the angle of friction, however this yields a non-determinate situation for analysis. For our initial foray into this theory we make the simple zero-slope assumption. This is conservative as it provides a maximal distribution of the displaced material across the surface. Redistribution of the zero slope material to have higher slope angles will in general create larger radius peaks and introduce, potentially, additional discontinuities within the distribution, as seen in \cite{minton}. Furthermore, this allows us to uniquely solve for and balance the redistribution of volume as the surface evolves. We note that in Harris et al.\ \cite{harris_tide} they use a more realistic model for regolith slope distributions, but do not consider the evolution of a given shape as it is spun up. Rather, their algorithm searches for a shape at a given angle of repose that matches the shape of 1999 KW4 Alpha. 

Ultimately, the fate and shape of the redistributed material is somewhat moot, for two reasons. First, the displaced material will be seen to uniformly lie above the Roche Lobe, and thus in the absence of energy dissipation in the flow would be able to escape from the body. Second, under an assumption that immediate escape does not occur, the radius profile of the redistributed regolith exceeds the limit for $a_x = 0$, and thus can be shed from the body and into orbit. Our assumption, again to be addressed with more detail in the future, is that material released from the asteroid will then be subject to escape. 

\begin{figure}[h!]
\centering
\includegraphics[scale=0.35]{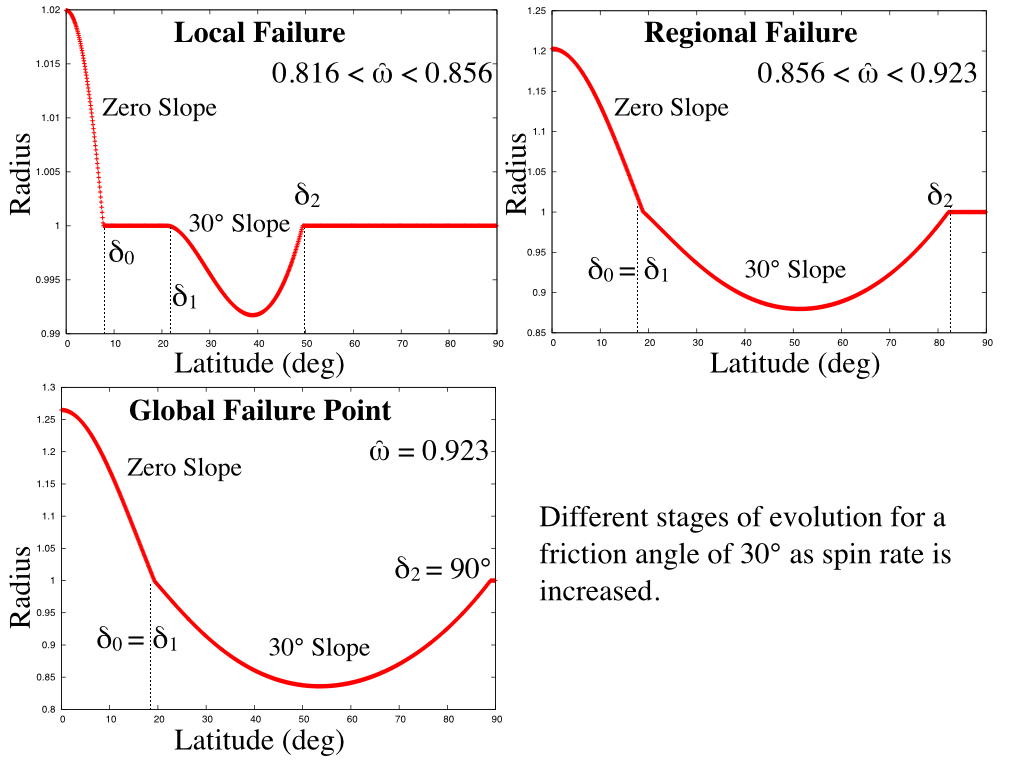}
\caption{Graphical depiction of the different phases of failure. }
\label{fig:regimes}
\end{figure}

In the following subsections we map out different regimes for our model as the spin rate is increased. We call these the Localized, Region and Global Failure Regimes, shown in Fig.\ \ref{fig:regimes}. 

\subsection{Initiation of Failure}

For spin rates such that the slope angle is less than the friction angle over the entire surface, $\theta < \phi$, then the regolith on the surface will not fail. To evaluate when this condition is first violated, recall that the maximum slope across the surface at a given $\tilde\omega$ equals $\tan\theta^* = \tilde\omega^2 / 2\sqrt{1-\tilde\omega^2}$. Evaluating the condition for when $\tan\theta^*=\tan\phi$ generates the condition
\beq
	\tan\phi & = & \frac{ \tilde\omega^2}{2\sqrt{1-\tilde\omega^2}} 
\eeq
This can be solved for $\tilde\omega$ to find
\beq
	\tilde\omega & = & \sqrt{ \frac{2\sin\phi}{1+\sin\phi}} \label{eq:fail}
\eeq
We note that the usual range of friction angles for granular media vary between 30 and 45 degrees. Across this range we see that the spin rates for failure range from 0.8 to 0.9 of the surface disruption spin rate, in agreement with the general observations of \cite{harris_tide}. 


Harris et al.\ \cite{harris_tide} noted that this condition was similar to that found by Holsapple for the global failure of a cohesionless ellipsoid \cite{holsapple_original}. 
We note that the condition in Eqn.\ \ref{eq:fail} is precisely the same limit at which a sphere with a uniform friction angle will begin to undergo global deformations when analyzed using a cohesionless Mohr-Coloumb failure law \cite{holsapple_original}. The equivalence of these two failure criteria, even though analyzed using very different approaches, is significant and indicates a level of consistency between these models. We note that our nominal model is of a relatively shallow mantle of regolith covering a rigid sphere, as modeled in \cite{hirabayashi_bilayer}. For this configuration global failure is delayed relative to a sphere consisting wholly of cohesionless regolith. Thus we do not consider global failure of the entire body. 

\subsection{Localized Failure Regime}

Once failure begins, our model localizes it to separate regions of the asteroid surface. 
The redistribution results in a small region of constant zero slope from the equator to a latitude denoted as $\delta_0$, and a region of constant slope $\phi$ between latitudes $\delta_1$ and $\delta_2$. These two regions will be disjoint from each other. We first compute the extent of the failure zone and its associated volume, and then evaluate the limits of the redistributed regolith. Since the region being cut-off will naturally have slopes greater than $\phi$, this means that a chord of constant slope $\phi$ must excise regions at higher latitudes beyond where the slope exceeds $\phi$. This first stage of failure ends when the transported regolith from the failure zone has a large enough volume to fill a region extending from the equator at a constant height up to and beyond the latitude at which the slope exceeds the friction angle. 

In this regime we must find the latitude at which the surface slope exceeds the angle of friction. Assuming that $\tilde\omega$ is greater than the limit in Eqn.\ \ref{eq:fail}, equating $\tan\theta = \tan\phi$ in Eqn.\ \ref{eq:slope2} yields
\beq
	\tan\phi & = & \frac{\tilde\omega^2\tan\delta}{(1-\tilde\omega^2)+\tan^2\delta}
\eeq
This equation defines a quadratic equation in $\tan\delta$ which can be solved to find the points where the surface first exceeds the angle of friction and where it stops exceeding the angle of friction.  
\beq
	\tan\delta & = & \frac{\tilde\omega^2}{2}\cot\phi \left[ 1 \pm \sqrt{1 - \frac{4(1-\tilde\omega^2)}{\tilde\omega^4}\tan^2\phi} \right]
\eeq
The ``-'' sign corresponds to the initial latitude when failure occurs and is denoted as $\delta_1$. 

If the regolith above this latitude assumes a constant angle $\phi$, the radius of the surface will conform to Eqn.\ \ref{eq:radius_slope}
\beq
	r_{\phi}(\delta,\delta_1) & = & \sqrt{\frac{ 1-\tilde\omega^2\cos^2\delta_1}{ 1-\tilde\omega^2\cos^2\delta}} e^{(\delta-\delta_1)\tan\phi}
\eeq
where $r_{\phi}(\delta_1,\delta_1) = 1$, and $r_{\phi}<1$ for values of latitude increasing from $\delta_1$. The failure zone will continue at least up to the higher latitude where $\tan\theta < \tan\phi$ again. However, since the surface slope is higher than $\phi$ up to this point, the region of constant slope must extend beyond this and up to a latitude $\delta_2$, defined by the condition $r_\phi(\delta_2,\delta_1) = 1$, for $\delta_2 > \delta_1$. This is a transcendental equation for $\delta_2$ and thus must be solved numerically. 

Once the limits of the local failure are found, $\delta_1$ and $\delta_2$, the volume displaced between these limits can be computed as $\Delta V_{\phi}(\delta_2,\delta_1)$ by combining Eqns.\ \ref{eq:vsurf} and \ref{eq:vol_const} in Eqn.\ \ref{eq:deltaphi} 
\beq
	\Delta V_{12} & = & \frac{1}{3} \Delta\lambda \left[ \left( \sin\delta_2 - \sin\delta_1\right) - \int_{\delta_1}^{\delta_2} \left( \frac{r_\phi}{R} \right)^3 \cos\delta \ d\delta \right] \label{eq:deltaV12}
\eeq

This volume must be equated to an equivalent volume of material deposited at the equator at a zero slope condition, with a radius that is larger than $1$, and is denoted as $1+H$. The radius profile of this zero slope region is found from Eqn.\ \ref{eq:radius_slope} by setting the slope angle to zero and the nominal radius to $1+H$. 
\beq
	r_H(\delta) & = & (1+H) \sqrt{\frac{1-\tilde\omega^2}{1-\tilde\omega^2\cos^2\delta}}
\eeq
Then  the corresponding volume under this curve is found by evaluating the integral in Eqn.\ \ref{eq:volume}
\beq
	V & = & \frac{1}{3} (1+H)^3 \Delta\lambda \sqrt{\frac{1-\tilde\omega^2}{1-\tilde\omega^2\cos^2\delta}} \sin\delta
\eeq
and the volume of material above the initial spherical shape is then
\beq
	\Delta V_H & = & \frac{1}{3} \Delta\lambda \left[ (1+H)^3 \sqrt{\frac{1-\tilde\omega^2}{1-\tilde\omega^2\cos^2\delta}} -1 \right] \sin\delta
\eeq

This volume of deposited material will extend up to latitude $\delta_0$, defined by $r_H(\delta_0) = 1$, leading to the special relation
\beq
	1+H & = & \sqrt{\frac{1-\tilde\omega^2\cos^2\delta_0}{1-\tilde\omega^2}} \label{eq:h}
\eeq
which, when substituted into the expression for $\Delta V_H$ yields the simplified result
\beq
	\Delta V_H & = & \frac{1}{3} \Delta\lambda \frac{\tilde\omega^2}{1-\tilde\omega^2} \sin^3\delta_0
\eeq

Finally, equating the terms $\Delta V_H(\delta_0) = \Delta V_{12}$ allows us to solve for the latitude at which the zero slope region of redeposited regolith will extend to. Specifically, it is found as
\beq
	\sin^3\delta_0 & = & \frac{1-\tilde\omega^2}{\tilde\omega^2} \left[ \left( \sin\delta_2 - \sin\delta_1\right) - \int_{\delta_1}^{\delta_2} r_\phi^3 \cos\delta \ d\delta \right] 
\eeq
Solving for $\delta_0$, the altitude $H$ is then found from Eqn.\ \ref{eq:h}. 

\subsection{Regional Failure Regime}

The localized failure regime is defined as long as $\delta_0 < \delta_1$. When this is violated, then the redistributed regolith and the slope failure regions run into each other, and defines a new failure regime we call ``regional failure.'' 
In regional failure there is a balance between the constant altitude covering from the equator and the constant slope region that still reintersects the surface before the polar region. To compute this balance the latitude $\delta_1$ where the transition between constant altitude and constant slope with angle $\phi$ occurs must be found, and the angle $\delta_2$ where this constant slope intersects the surface again. At latitudes higher than $\delta_2$ the surface will still be less than the angle of repose and would not have failed yet. 

When in the regional failure regime, the disturbed regolith will flow from the upper latitude $\delta_2$ down to an intermediate latitude at a constant slope angle of $\phi$, where it will directly intersect with the redistributed regolith at zero angle of repose. Technically, this means that $\delta_0 = \delta_1$ in the above analysis. The defining conditions are that the displaced volume balance still holds and that there exists an upper latitude $\delta_2$ where the surface reverts back to its original spherical slope. The conditions can be restated as
\beq
	\Delta V_H(\delta_1) & = & \Delta V_{12}(\delta_2,\delta_1) \\
	r_{\phi}(\delta_2,\delta_1) & = & 1
\eeq
Both equations are transcendental relations and must be solved for numerically. We have found that a simple iteration process will reliably converge on the solution. For the given value of $\tilde\omega$, first solve for $\delta_1$, $\delta_2$ and then $\delta_0$, following the local failure algorithm given in the previous subsection. Then substitute $\delta_0$ for $\delta_1$ in the process, which enables one to find $\delta_2$ and, again, an updated value of $\delta_0$. This new value is used again, with the iteration stopping when the change in $\delta_0$ is smaller than some threshold. We have found this algorithm to converge for all relevant conditions tested. 

The latitude at which this transition between constant slope and zero slope occurs does not change much as the spin rate is increased, for most cases by just a few degrees. If it were assumed that the redistributed material at the equator had a higher slope, then the transition to regional failure would be delayed. In this case it is possible to pile the redistributed regolith into a pile with a larger radius at the equator, or even redistribute the material in a more complex way. In this sense, by assuming a zero slope redistribution the profiles we find are conservative. 

The regional failure regime lasts until the upper latitude $\delta_2$ reaches the pole. Once this occurs all surface material has failed or has been covered, and is able to redistribute itself across the body surface. We denote that this condition as ``global failure'' and note that it occurs at values of $\tilde\omega < 1$.

\subsection{Transition Points and Limits on Radius}

For this simple model the transition points for a spinning sphere as a function of angle of friction can be determined. As $\phi$ is varied the following can be tracked: the spin rate at which failure starts and the latitude at which is starts, the spin rate and upper and lower latitudes at regional failure, and the spin rate and transition latitude at global failure. Table \ref{tab:1} marks these numerically determined transition points. 

\begin{table}[h!]
\centering
\begin{tabular}{|| c || c | c || c | c | c || c | c ||}
\hline
\hline
& \multicolumn{2}{c||}{First Failure} & \multicolumn{3}{c||}{Regional Failure} & \multicolumn{2}{c||}{Global Failure} \\
$\phi$ & $\tilde\omega$ & $\delta_1$ & $\tilde\omega$ & $\delta_1$ & $\delta_2$ & $\tilde\omega$ & $\delta_1$ \\
(deg) & & (deg) & & (deg) & (deg) & & (deg) \\
\hline
\hline
30 & 0.816 & 30.0 & 0.856 & 14.99 & 65.34 & 0.923 & 19.24 \\
\hline
35 & 0.854 & 27.5 & 0.887 & 13.77 & 60.68 & 0.954 & 17.44 \\
\hline
40 & 0.885 & 25 & 0.911 & 12.50 & 55.18 & 0.976 & 15.19 \\
\hline
45 & 0.910 & 22.5 & 0.932 & 11.27 & 50.68 & 0.989 & 12.59 \\
\hline
50 & 0.931 & 20.0 & 0.948 & 9.92 & 45.00 & 0.996 & 9.65 \\
\hline
\hline
\end{tabular}
\caption{Values of spin rate at which transitions occur for different values of friction angle. The stated latitude are, respectively, the latitude at which failure first occurs, at which the constant slope region first meets the zero slope region and the upper latitude at this point, and the lower latitude when these regions reach the global failure condition.}
\label{tab:1}
\end{table}

\begin{figure}[h!]
\centering
\includegraphics[scale=0.25]{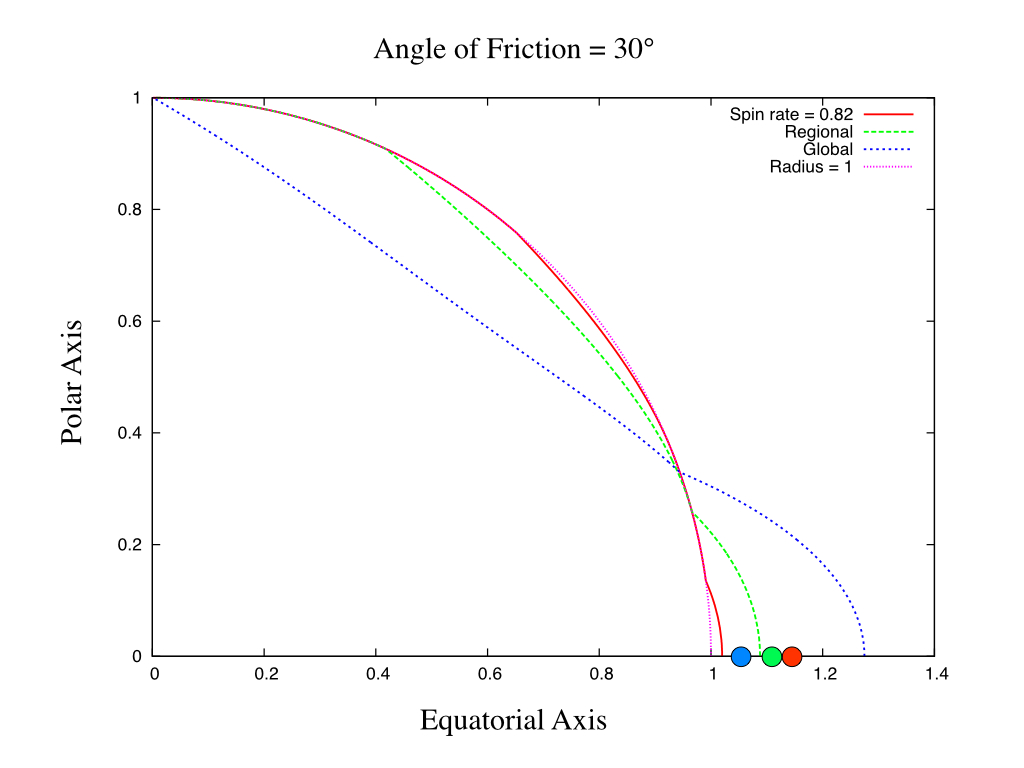} 
%
\includegraphics[scale=0.25]{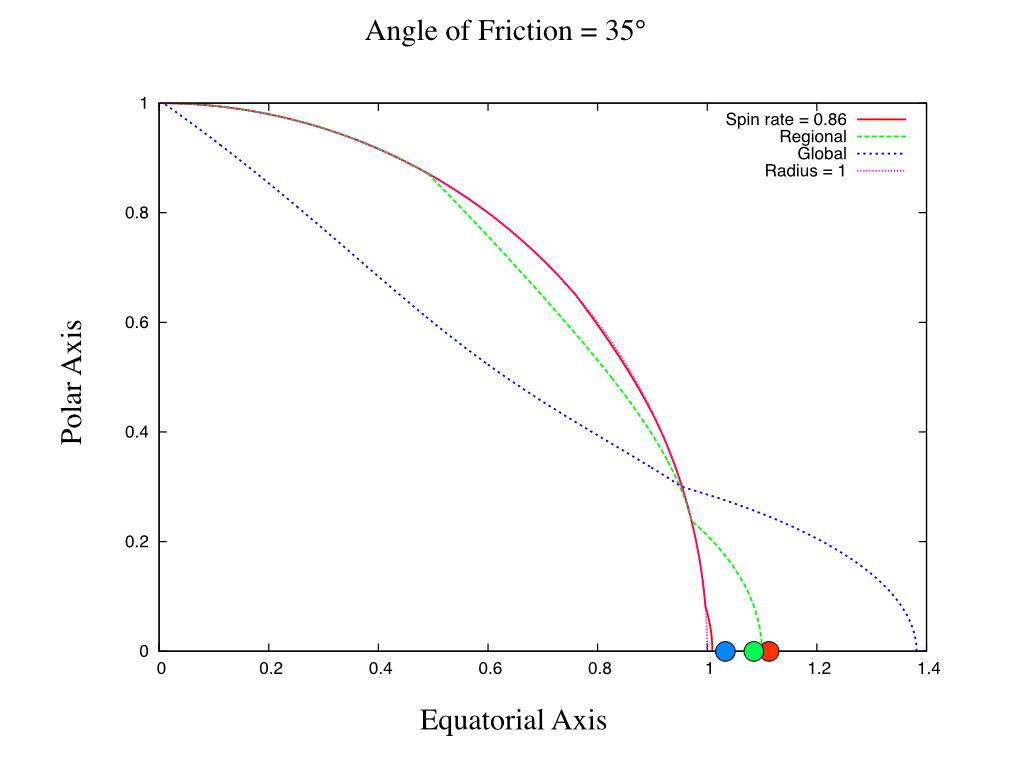} 
\caption{Radius plots for a range of spin rates at significant transitions for a friction angle of 30$^\circ$ (top) and 35$^\circ$ (bottom). The axes are normalized distance. The dots show the location of the equilibrium point (i.e., the intersection of the Roche Lobe with the equator) at the different spin rates shown.}
\label{fig:rad35}
\end{figure}

In Fig.\ \ref{fig:rad35} we show these ideal radius profiles for values of friction angle of $30^\circ$ and $35^\circ$ at their transition spin rates. These are now plotted in a polar graph, showing the explicit shape of the hemisphere. Also shown is the corresponding location of the equilibrium point at each of the significant transition points (the colored dots along the horizontal axis). It is significant to note that at $35^\circ$ the limiting radius of the shape profiles lies within the redistributed regolith radius profile, meaning that a significant portion of the redistributed material is beyond the equilibrium point and will be lost from the surface. At higher friction angles this intersection occurs well before the regional failure limit, indicating that the equatorial radius profile should be limited in extent by the equilibrium point radius for the largest spin rate the body would have experienced. Thus, the large deformations in radius seen in these plots are most likely not present, having been shed from the body. 

%
%

\begin{figure}[h!]
\centering
\includegraphics[scale=0.25]{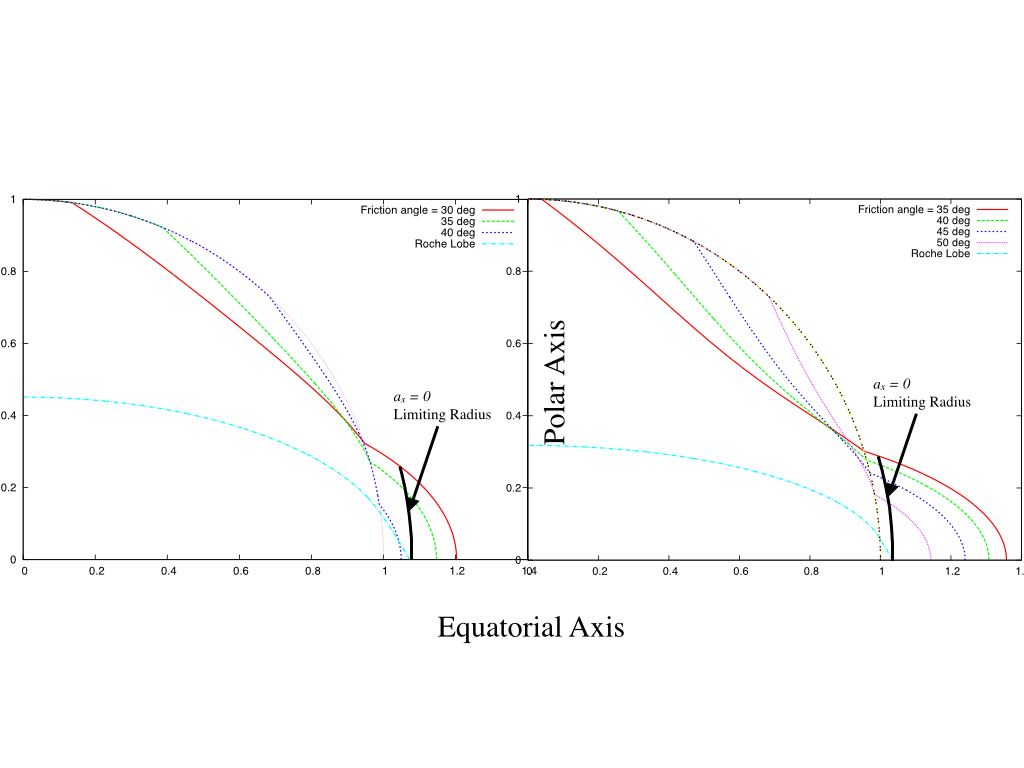} 
\caption{Radius plots for a range of friction angles at a spin rate of $\tilde\omega = 0.90$ (left) and 0.95 (right). The axes are normalized distance. }
\label{fig:ZV_090}
\end{figure}

The implications of this is that when the regolith on a spinning body passes its angle of repose and flows, it should be common that some of this material is shed and leaves the region about the asteroid surface and can then escape from the system. Figure \ref{fig:ZV_090} shows a range of surface profiles at two different spin rates for a range of friction angles. Also shown is the Roche Lobe radius, extending from the interior to the exterior. From these figures it is clear that all of the displaced material starts out above the Roche Lobe, and thus unless energy is dissipated while the grains flow, this material is able to escape directly. If it is redistributed, however, it is clear that the bulge will be truncated, with the degree of truncation controlled by how rapidly the asteroid has spun in the past. Specifically, the more rapid the past spin of the body has been, the less pronounced its equatorial bulge should be, when accounting for the current model alone. 

\subsection{Rates of Surface Shedding}

\begin{figure}[h!]
\centering
\includegraphics[scale=0.25]{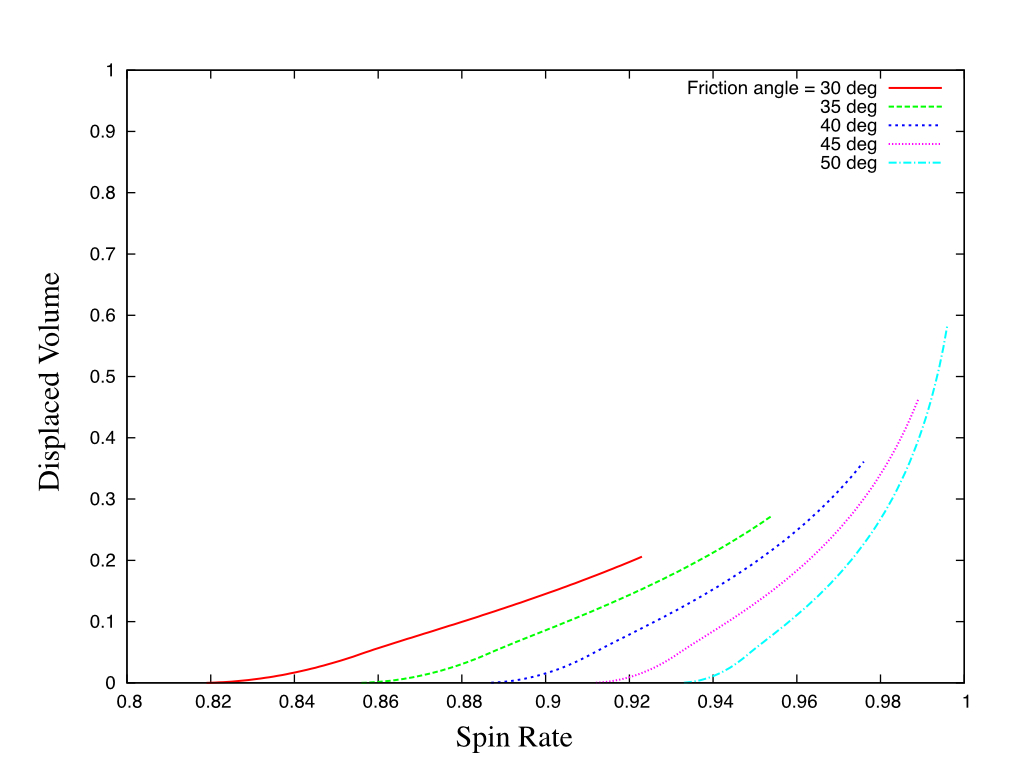} 
\caption{Normalized volume displaced and possibly lost as a function of spin rate and friction angle. Curves are stopped once the global failure limit is reached.}
\label{fig:volumes}
\end{figure}

\begin{figure}[h!]
\centering
\includegraphics[scale=0.25]{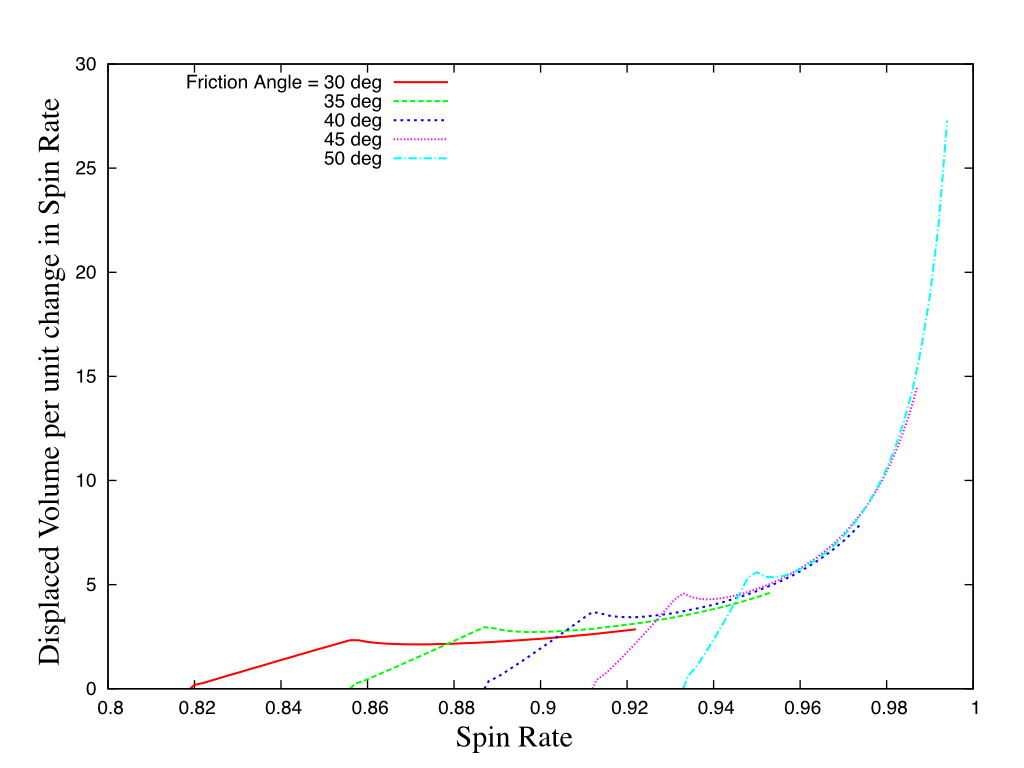} 
\caption{Sensitivity of volume growth with respect to changes in normalized spin. This plot consists of the slopes of Fig.\ \ref{fig:volumes} as a function of normalized spin. Note the transition in slopes as the failure goes from local to regional.}
\label{fig:vslope}
\end{figure}

If such a loss of material occurs, estimates can be given for the total volume of material sent into orbit. An upper bound on the mass lost can be tracked by computing the total volume of displaced material as a function of spin rate and angle of friction. The relevant volume integral is defined in Eqn.\ \ref{eq:deltaV12}, and is computed for both local and region failure limits. This computation is shown in Fig.\ \ref{fig:volumes} as a function of spin rate for different friction angles. Although the total amount of volume loss can become very large, this loss would only occur incrementally as the spin rate of the object is increased. Figure \ref{fig:vslope} shows the derivative of the curves in Fig.\ \ref{fig:volumes}, specifically $\frac{\partial\Delta V}{\partial\tilde\omega}$, which can be used to estimate the rate at which volume is released given an constant increase in spin rate.

If a body is rotationally accelerated to the point where surface failure occurs, and if the body continues to be rotationally accelerated at a rate $\dot{\tilde{\omega}}$, then the average rate at which its volume is lost is simply
\beq
	\Delta \dot{V} & = & \frac{\partial\Delta V}{\partial\tilde\omega} \dot{\tilde\omega}
\eeq
where the sensitivity can be read off of Fig.\ \ref{fig:vslope} as a function of friction angle and $\tilde\omega$. Estimates for $\dot{\tilde\omega}$ can be found by applying YORP theory to the body in question, as this is the prime source of secular increases in spin rate of small asteroids. A specific application of this result is given in the Discussion section. 

\clearpage

\section{Discussion}

In the following we explore some implications of this theory and apply our analysis to some realistic asteroid models and situations. 
First, the current theory is placed into context in relation to other types of rubble pile body failures. Then the shape-related observables are defined and compared with existing spheroidal asteroid models. Following this, the implications of this theory for surface shedding are reviewed and compared to the recently observed active asteroid P/2013 P5. Finally, significant assumptions made in the theory are reviewed along with a brief discussion on what their likely impacts may be. 

\subsection{Fission and Surface Shedding}

The sort of asteroid failure described in this paper has similarities with the model proposed by Walsh \cite{walsh_nature} but is distinct from other models of failure that have been proposed and, indirectly, observed in the asteroid population. This analysis is specifically focused on the evolution of spheroidal bodies with rotational symmetry, the outcome of this initial assumption feeds the analysis throughout.
This is to be distinguished from the failure or fission of bodies modeled either as contact binaries or as elongate rubble piles, which have been studied both theoretically and numerically in detail over the past decade (c.f., \cite{scheeres_icarus_fission, sanchez_icarus, jacobson_icarus}). In those situations analysis has shown that failure frequently occurs via the splitting of the body into multiple components, and significantly, observations of asteroid pairs have been directly linked to this mode of rotational fission \cite{pravec_fission, polishook_pairs}. 
We specifically note that this rotational fission failure mode is markedly different than the effect explored here. 

\subsection{Observable Features in Asteroid Shapes}

Our theory predicts a number of specific features that should be observable in the shapes of asteroids that have undergone such surface failures. Although ground based models of spheroidal asteroids are not precise enough to unambiguously support the geophysical interpretation of these features, in the near future two missions will be visiting spheroidal asteroids and ideally taking precise enough measurements to help identify whether such features exist. These are the Hayabusa 2 mission to asteroid  (162173) 1999 JU3 and the OSIRIS-REx mission to asteroid (101955) Bennu. 

\paragraph{Application to Actual Asteroid Shapes}
Despite the limitations of the current asteroid models, we can still discuss and explore how observational features could be applied to existing asteroid shape models. Indeed, the general trends found for the simple spherical systems carry over to the realistic asteroid shapes. For the shapes, the most striking differences are due to the lack of symmetry and to variations in the object's geopotential. The major change which occurs is that the gravitational potential no longer has a simple form, although it can be expressed in terms of the total mass as $U(\bfm{r}) = {\cal G} M \tilde{U}(\bfm{r})$, where $M$ is the total mass and $\tilde{U}$ is the unit mass potential. The unit mass potential can contain both geometric variations and density variations, and can be determined from existing polygonal models (c.f., \cite{scheeres_asteroid_book}). This allows one to compute the same effective parameter $\tilde\omega$ for these more complex shapes, equal to $\omega \sqrt{\frac{3}{4\pi {\cal G} \rho}}$, where $\rho$ is the bulk density of the body. It should be noted that the values of this parameter where significant changes occur on these bodies are not preserved from the spherical case. Due to non-spherical shapes, it is found that higher slopes can be achieved for lower values of $\tilde\omega$, in general. 

\begin{figure}[h!]
\centering
\includegraphics[scale=0.1]{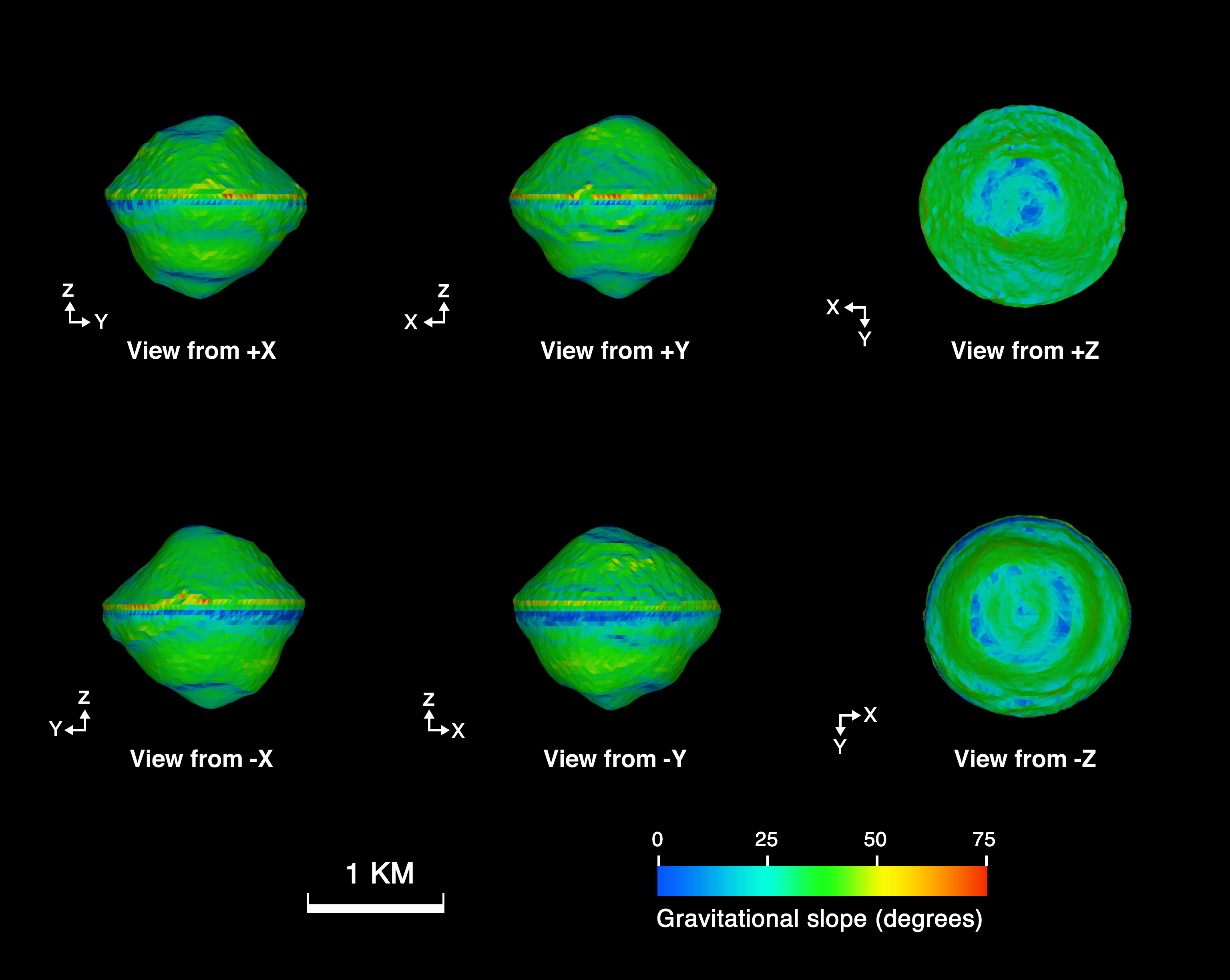}
\caption{Surface slopes on 1999 KW4 Alpha \cite{KW4_ostro}.}
\label{fig:KW4_slope}
\end{figure}

\begin{figure}[h!]
\centering
\includegraphics[scale=0.3]{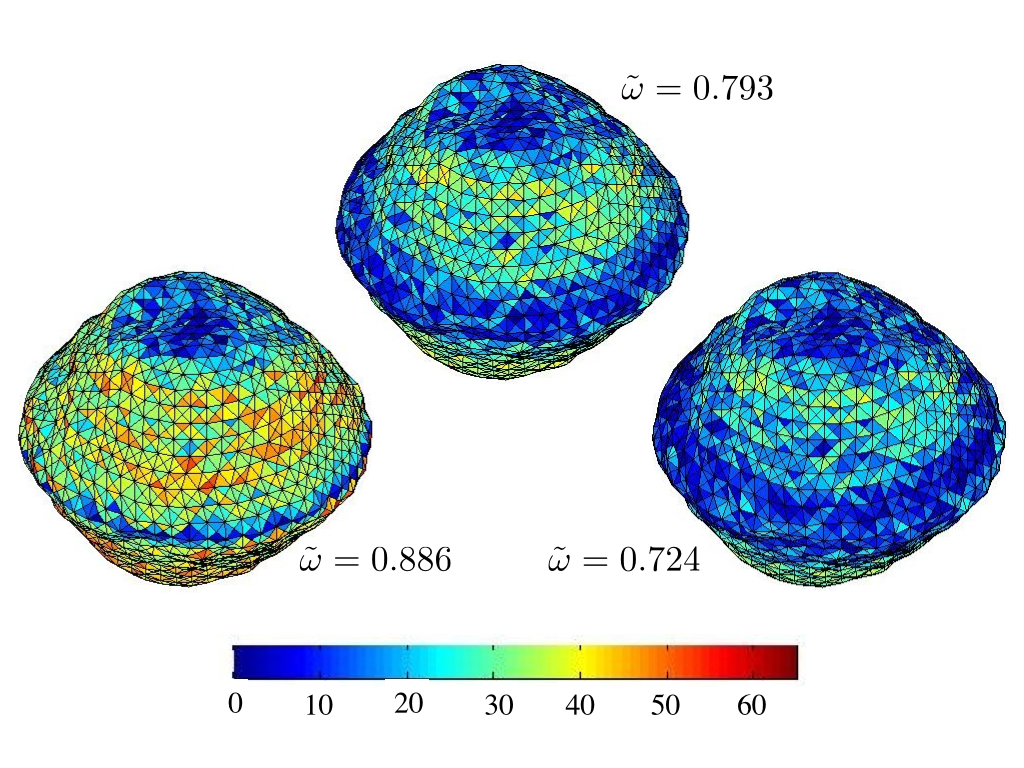}
\caption{Surface slopes on 2008 EV5 for different values of $\tilde\omega$.}
\label{fig:2008EV5}
\end{figure}

We focus on two asteroids, 1999 KW4 Alpha \cite{KW4_ostro} and 2008 EV5 \cite{busch_EV5} (Figs.\ \ref{fig:KW4_slope} and \ref{fig:2008EV5}, respectively). Both have the characteristic spheroidal shape which this paper is concerned with. We find that 1999 KW4 fits a general description of a body that has had significant surface failure and which is currently close to this limit. Based on radar measurements \cite{KW4_ostro} the density of the primary is $\sim 2$ g/cm$^3$ and it has a spin rate of 2.764 hours. These combine to yield a $\tilde\omega = 0.83$ for this body. 
The asteroid 2008 EV5 has shape features which could be consistent with past surface failure, although whether it is currently at these limits is not known as there are no firm estimates of its density. It is important to note that current asteroid models for bodies near these disruption conditions are not precise enough to clearly distinguish what is going on. This theory, however, provides constraints which allow us to ascertain what fundamental geophysics may be at play at a rotationally symmetric body such as found for binary asteroid primaries like 1999 KW4 or at single bodies like 2008 EV5. 

In the next few paragraphs we explicitly consider the shape and slope based observable features from our theory that could be visible in these real asteroid shapes. We note that both of these shapes clearly have a strong North/South symmetry in their shape and slope properties. 

\paragraph{Uniform Slopes over Mid-latitudes}
Once surface failure occurs the material that comes off of the mid-latitudes should leave behind material at the angle of repose for the regolith. Harris et al. \cite{harris_tide} noted that it is expected that there will be a large range of latitudes at a constant level of surface slope. It is important to note that, due to further spin rate evolution of a given asteroid that its current surface need not currently be at a uniform slope value. In fact, should the spin rate of a spheroidal asteroid decrease from its maximum rate, such as via YORP deceleration, then the shape of the body would remain fixed even though the slopes over the mid-latitudes would uniformly decrease. Conversely, the surface of a body that undergoes a continual spin-up in rate would continue to evolve its surface, exposing a growing region at the angle of repose while shedding material. Due to this, to test whether a given spheroidal body has experienced such a limiting behavior in the past it is necessary to evaluate the model across a range of increased spin rates to see if, at a specific rate, the body's mid-latitudes take on a uniform slope. The presence of such a feature would indicate what the angle of repose of the body's regolith is. We apply this idea directly to two asteroids in the following section. 

First consider 1999 KW4 Alpha, seen in Fig.\ \ref{fig:KW4_slope} and also as described in \cite{harris_tide}. It is clear that the mid-latitudes of this body are in the vicinity of a constant angle of $\sim 40^\circ$, and that these slope values should be relatively accurate as the total mass and volume of this body have been estimated based on radar and astrometric observations. This is consistent with the theory and indicates that the surface regolith of this body has an angle of repose around this value. Of course, the shape asymmetries lead to some significant variations of the local slope, and without more detail we cannot tell whether these are due to unique topographical features or represent a range of angle of repose values in that asteroid's regolith. Still, the uniformity of the slopes are striking. 

\begin{figure}[h!]
\centering
\includegraphics[scale=0.3]{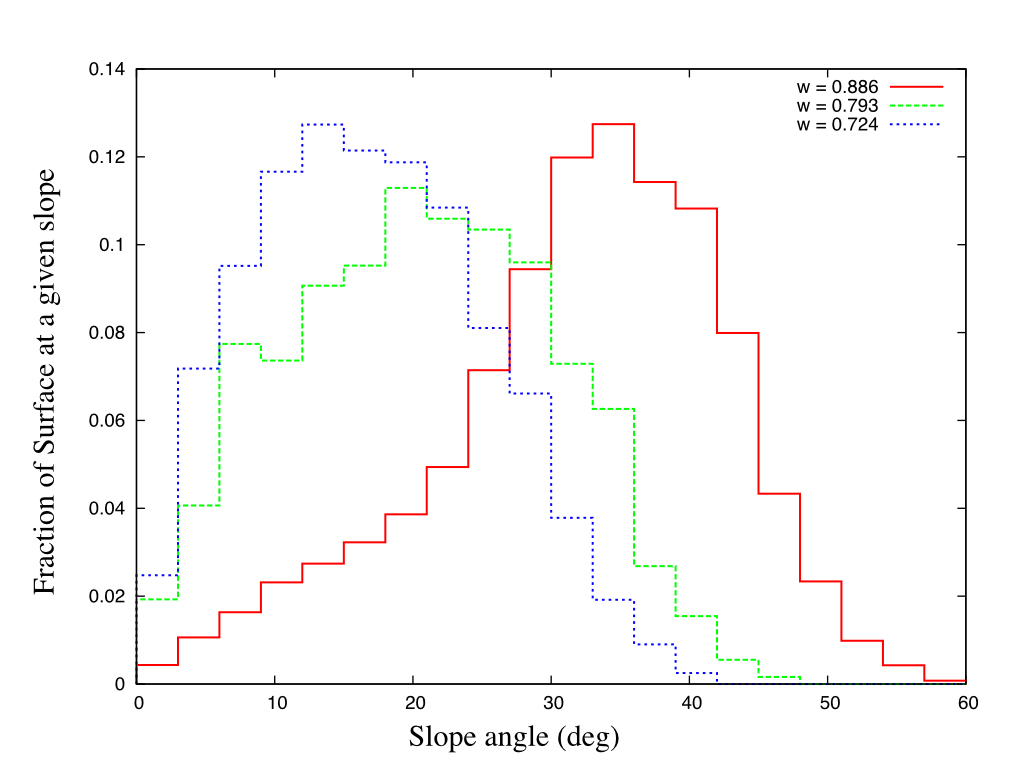}
\caption{Slope histograms on 2008 EV5 for different values of $\tilde\omega$. Note that the slope distribution narrows somewhat for the fast spin rate. }
\label{fig:slopeEV5}
\end{figure}

Regarding 2008 EV5, as discussed above, it is important to note that the actual shape of a spheroidal asteroid's mid-latitudes may be more informative than its current slope value, as it is quite feasible that the spin rate of a given body has undergone a rotational deceleration due to YORP from the time of its failure or that we do not have an accurate current density estimate for this body. 
One approach to probe this is to recompute the surface slopes for this body as the density and spin rate is varied (i.e., as a function of $\tilde\omega$) is to search for spin rates at which the mid-latitudes take on a more uniform slope structure, yet material is not shed from the equator. Such an approach can be used on any body with an accurate shape model as a means to estimate the angle of repose of regolith on that body, independent of having a good density estimate. Applying this approach to 2008 EV5, shown in Figs.\ \ref{fig:2008EV5} and \ref{fig:slopeEV5} we find that at a normalized spin rate of 0.886 the mid-latitudes of this body take on a relatively uniform slope structure between $35-45^\circ$. It is also appropriate to note that this is just shy of the spin rate at which the equator experiences zero acceleration, discussed later. We must note, however, that this approach is not definitive, as one may also be able to interpret the slope structure at lower spin rates as being consistent with a lower angle of repose. At this level of detail it becomes important to provide a focused and rigorous analysis of the given shape, something which is beyond the current paper's scope. 

\paragraph{Transitions in Slope at Upper Latitudes} 
A related feature of our failure model is that the regions of constant slope may not extend all the way to the polar regions. Only if the body has been spun past its global failure limit are the slopes around the poles non-zero. Otherwise, the polar regions should have slopes that decrease towards smaller values. Once a detailed shape model is in hand, it may even be possible to further constrain the failure conditions and material angle of friction by comparing these transition latitudes with detailed models. 
To analyze this feature there is a need for detailed shape and slope maps of the higher latitude regions. It should be noted that radar observations often do not fully sample these regions. 

We see that both of our asteroids exhibit some aspects of this feature. 2008 EV5 clearly shows a transition from higher-slope to a low-slope cap across the poles of that body. For 1999 KW4 this feature is more subtle, and while a break in slope values is clearly seen, we also note that the polar regions do not have a smooth slope appearance. This could either indicate global failure or a different, and perhaps complementary, mode of failure of this body if the gravitational pressure along the rotation axis of this body overcomes the strength of material across the equatorial plane, as described in \cite{hirabayashi_LPSC14}. To better understand such possibly combined features it is necessary to carry out more detailed and geophysically consistent analyses of the evolution and failure states of such a body. 

\paragraph{Radial Limits of the Shape} 
Another observable feature that our model predicts is that a band about the equator of the spheroidal asteroid should be limited by the radius of the fastest equilibrium point that it would have experienced in the past. We see that, ideally, this would actually impose a nearly spherical arc about the equator with non-zero slopes, which would then (in our idealized model) transition to a zero-slope regime and then a constant slope regime (see Figs.\ \ref{fig:ZV_090} and \ref{fig:ZV_095}). Of course, the detailed morphology of the slopes in the vicinity of the cut-off in radius is expected to be more complex, with our model just providing a simple and possible end-state. Again, as the current spin state of the asteroid is not necessarily at this failure limit, one must determine how fast the body must spin before the equatorial region again reaches this limit. It is also important to note that the actual spin limit at which material will be shed from the surface is very sensitive to variations in the body's density distribution, and thus may not fully conform to a simple constant density model of the body. 

We first note that for 2008 EV5, that at a spin rate slightly higher than the 0.886 value shown, the equator of that body was subject to outward accelerations, indicating a lack of ability to retain material. Thus we see that the spin rate at which the surface slopes uniformly occupy most of the mid-latitude regions corresponds to the rate at which the surface of the body becomes disrupted. This is certainly consistent with our current theory, and can in fact place a lower limit on the density of this particular asteroid. 2008 EV5 has a spin period of 3.725 hours \cite{busch_EV5}, and thus a normalized spin rate of 0.886 is equivalent to a density of $\sim 1$ g/cm$^3$, consistent with other primitive asteroids \cite{chesleybennu}. 

\begin{figure}[h!]
\centering
\includegraphics[scale=0.25]{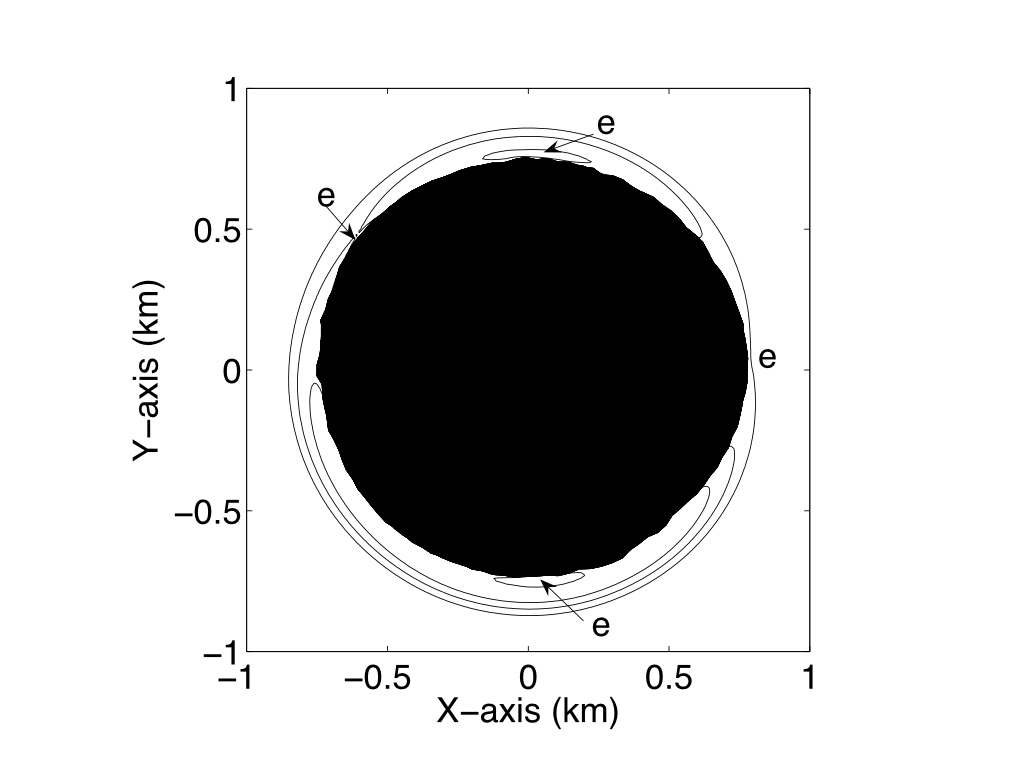}
\caption{Top-down view of the Roche Lobe and equilibrium points of 1999 KW4 (figure from \cite{KW4_scheeres}). Note that close proximity of the equilibrium points to the surface, indicating that this body may be at or close to its current surface failure spin rate. }
\label{fig:1999KW4eq}
\end{figure}

\begin{figure}[h!]
\centering
\includegraphics[scale=0.25]{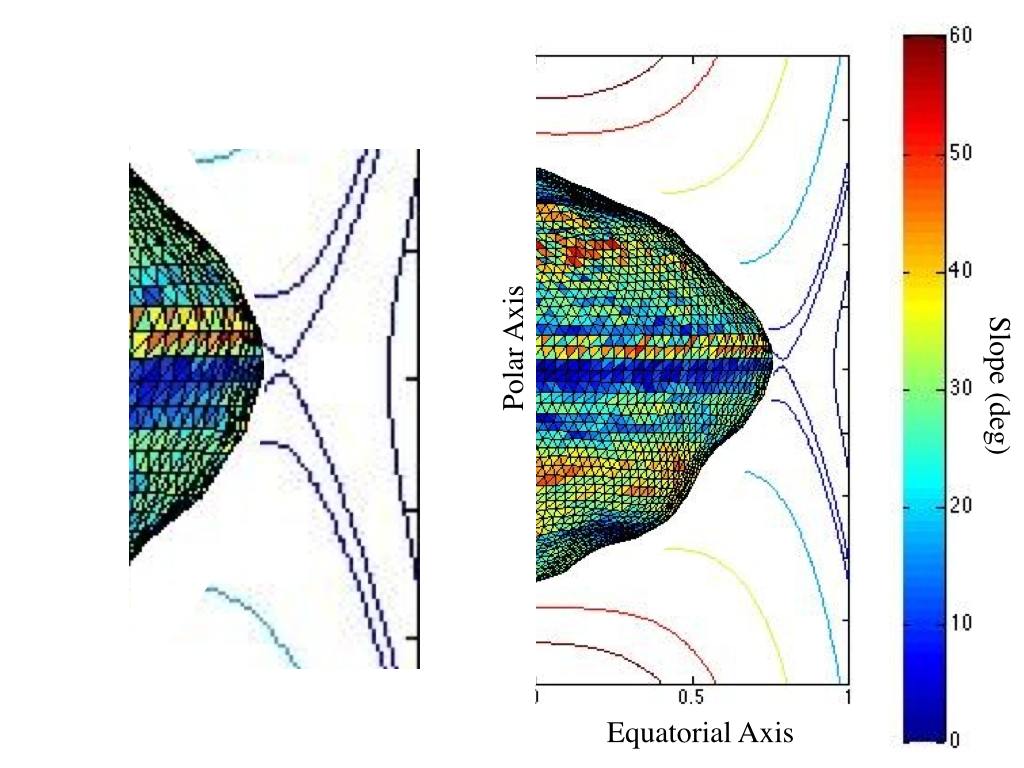}
\caption{Side view showing surface slopes on 1998 KW4 and Roche Lobe. Note that the figure is tuned so that the rightmost point on the body is shown in proper proximity to its Roche Lobe. The detail on the left shows a characteristic constant radius segment of the shape around the equator. The lines are contours of equal geopotential. }
\label{fig:1999KW4}
\end{figure}

Comparison of the Roche Lobe of 1999 KW4 to the shape of the asteroid provides a real striking example of this effect. Figures \ref{fig:1999KW4eq} and \ref{fig:1999KW4} show top-down and  side views of the lines of constant geopotential for this asteroid.  
From these plots we see that the equilibrium points are very close to the asteroid surface, and indeed small variations in the estimated density (within the error bounds) could even bring the equilibrium points directly to the surface of the body. The theory predicts that the lofted regions of the surface should consist of a spherical band about the equation. The detail in Fig.\ \ref{fig:1999KW4} does not directly show this feature, as the equatorial profile has a tighter angle of curvature than a circle of constant radius centered on the asteroid. It should be noted, however, that the limiting radius of the asteroid will be sensitive to mass redistribution, not considered in the current theory. The specific shape may also be influenced by the dynamics of failed material, as comparison with the Roche Lobe indicates that the failed material could essentially roll off the surface and escape.
These specific limits must be probed in more detail, as noted later in this section.


\subsection{Mass Shedding Events}

The theory developed in this paper shows that surface landslides on asteroidal bodies naturally lead to transient mass shedding events of measurable volume. It is interesting to note that the theory indicates that if an object experiences a shedding event, and if the spin rate is continued to be accelerated, then there should be incremental mass shedding events in the future. This lends credence to the idea that asteroids can create meteoroid fields over a longer time span with multiple shedding events, and motivates future observations to determine the shape and morphology of the parent bodies of observed meteor streams. Such shedding events are also consistent with some aspects of the observed active asteroids. Once a surface is at the failure limit, any perturbation may cause instability in the surface slope and could lead to a loss of material. This could either be a modest impact event or potentially a close planetary flyby. In fact, depending on the timescales between YORP spin-up and impacts with small asteroids \cite{marzari_impacts}, it is reasonable to hypothesize that the dominant trigger for shedding events may be small impacts that locally destabilize surfaces close to failure. 

Our theory has implications and suggests possible interpretations of the recently observed active asteroids such as P/2013 P5 \cite{jewitt_P2013_A, hainaut_2013P5}, which was observed to have multiple streamers of grains escaping from it at low relative speeds (note, its cometary designation is due to initial observations that detected a coma-like structure, later shown to be asteroidal material). It is significant to note that there were several distinct shedding events observed over a period of time. Our current model tacitly assumes that these should all be related to each other, and that the asteroid may be undergoing a shedding ``phase'' that may have been triggered as described above. It is not unreasonable to assume that before shedding occurs the surface may be ÒperchedÓ in some aspects and be ready to fail at several different regions. Then, once some shedding occurs, the orbital environment about the body should be messy and may result in a period of secondary impacts and associated seismic shaking events which could generate additional losses of material, etc. However, in this model this period of shedding should not persist for a long time. 

The total mass in the observed dust distributions from microns to meters was estimated to be on the order of $2\times10^7$ kg \cite{hainaut_2013P5}, whereas the total mass of the asteroid can be estimated to be $\sim 2\times 10^{11}$ kg, if we assume that the parent body has a bulk density of 2 g/cm$^3$, commensurate with a porosity of at least 2/3. The observed material shed from the body is a small fraction, $10^{-4}$ of the total mass. 
While there is insufficient observational evidence to fully constrain the event using our current model, a simple comparison of the relative volume of material indicates that it was likely an incremental loss of material. This fits a model in which, as the object's spin rate is increased, the surface is pushed beyond its previous stable limit and additional material landslides and can be released from the surface. This material can then be lost, either by gaining sufficient kinetic energy to directly depart the vicinity of the asteroid, or by being pushed beyond the equilibrium radius where it would experience outward accelerations, separating the grains from the body. Of particular usefulness would be observations of the current spin rate of this body as such observations could determine if this object fits with our proposed model. 

We can actually calculate the possible average shedding rate of P/2013 P5 with a few assumptions. First assume that this body has been spun-up to and beyond the point where it has begun a phase of surface failure. If we assume that this is due to the YORP effect the angular acceleration can then be estimated as
\beq
	\dot\omega & = & \frac{G_1}{a^2\sqrt{1-e^2}} \frac{R}{M} {\cal C}
\eeq
using the notation from \cite{scheeres_YORP}, where $G_1 \sim 1\times 10^{14}$ kg km/s$^2$ is the solar constant, $a$ and $e$ are the semi-major axis and eccentricity of the asteroid's heliocentric orbit, $R$ is the mean radius of the asteroid, $M$ is its total mass, and ${\cal C}$ is its normalized YORP coefficient. In \cite{scheeres_YORP} it is found that the normalized YORP coefficient seems to vary between near-zero and 0.01 across a wide range of elongate asteroid shapes, while for the oblate shape of 1999 KW4 it was found to be at most 0.0004 \cite{CMDA_mirrahimi}. We do note that the YORP coefficient can be enhanced by the Tangential YORP effect \cite{golubov_TYORP}, although we do not consider that here. Replacing the mass with $M = 4\pi/3 \rho R^3$, and dividing by the disruption rate $\sqrt{4\pi/3 {\cal G} \rho}$ and evaluating the angular acceleration at 1 AU, we find
\beq
	\dot{\tilde\omega} & = & \frac{0.0635}{A^2 \sqrt{1-e^2}} \frac{{\cal C}}{\rho^{3/2} R^2}
\eeq
where $A$ is the semi-major axis in AU. For P/2013 P5 the following values apply, $A = 2.189$ AU, $e = 0.115$, and from \cite{hainaut_2013P5} $R = 290$ m and an assumed bulk density of $\rho = 2000$ kg/m$^3$. Then
\beq
	\dot{\tilde\omega} & = & 1.77 \times 10^{12} {\cal C} \mbox{ rad/s}  \\
	& = & 5.57\times10^{-5} {\cal C} \mbox{ rad/year}
\eeq

To convert this into a rate of volume loss, multiply $\dot{\tilde\omega}$ by $\partial \Delta V/\partial\tilde\omega$ from Fig.\ \ref{fig:vslope}. For a friction angle of $40^\circ$, we see that this sensitivity is less than 10. If we choose an intermediate value of $\partial \Delta V/\partial\tilde\omega \sim 5$ we find the fractional rate of mass loss per year as $\sim 2.8 \times 10^{-4} {\cal C}$. Then the length of time expected between mass loss events of the order of magnitude just seen for this body can be found by dividing the observed fractional mass loss by the above rate to find
\beq
	T_{loss} & \sim & \frac{0.36}{{\cal C}}
\eeq
Thus for a YORP coefficient similar to that of 1999 KW4, $\sim 0.0004$, this would predict this level of mass loss every $\sim 900$ years. The uncertainty in this estimate is large, a factor of several at least, and would require more information about the shedding body to better constrain. 

We note that this assumes that the YORP effect always acts in the positive direction, although it is well known that small variations in the shape of a body can have a large effect on the YORP coefficient \cite{scheeres_itokawa_YORP, statler}, potentially even making it go negative (although Tangential YORP should provide an overall positive bias \cite{golubov_TYORP}). Nonetheless, an extended period of positive YORP coefficient could drive such an asteroid to repeated shedding events. It is interesting to note that this also becomes a model for the creation of meteoroid streams from asteroids, and could modify recent interpretations of the persistence of asteroid-related streams, providing them a mechanism for sustained creation of debris as opposed to a model where debris is only generated from impacts \cite{iwan_williams}. 

\subsection{Exposure of Sub-Surface Material}

A final, potentially observable, aspect of this model is that once a landslide occurs, sub-surface material is exposed starting at mid-latitudes and then progressing across a wider region of the body. This opens interesting and exciting observational opportunities. Namely, if an object is currently in a shedding phase of its life it becomes possible to observe older surfaces that have been weathered and newer sub-surface material that has not undergone recent weathering. Specifically, this could lead to spectral heterogeneity across the surface of the asteroid, a feature which has been observed for some bodies. Detailed granular mechanics computations can trace the fate of surface vs. sub-surface material migration, and would be of specific interest for any mission that visits a spheroidal asteroid. For our simple model of failure, the surface material would tend to migrate from mid-latitude regions to the equator, along with a component of sub-surface material, exposing fresher material at mid-latitudes (if the transported surface material is not overturned). We note that the polar regions are predicted to remain largely unaffected by this migration, unless perhaps the body is spun up to the global failure limit. Such spin rates are quite extreme, however. 
We note that this is distinct from conclusions by Walsh et al. \cite{walsh_nature, walsh2012}, which found migration of blocks from the poles down to the equatorial region. Differences between those models and the current model may be related to the relatively low resolution in the Walsh et al. numerical models, or in the different physical evolution resulting from their use of a crystalline stacking to create high angles of friction. 

Recent spectral observations of some asteroids in the spheroidal class have clearly shown a large degree of spectral heterogeneity. A clear and significant example of this is asteroid 1996 FG3, which was subject to multiple observations at several epochs due to its prior interest as a rendezvous destination for space missions \cite{hussmann2012, tardivelacta, basixLPSC14}. The data on this body show that it has a spheroidal primary and that it is currently spinning at a rapid rate, with an estimated $\tilde\omega \sim 0.87$ \cite{wolters_FG3}. Several observation campaigns of the spectra of this body have clearly documented a surface that has significant spectral heterogeneity, which seem to be correlated to observation geometry -- meaning that the differences seem to be tied to location on the body \cite{deleon2011, deleon2013, rivkin2013}. A body undergoing surface shedding events is expected to expose fresh and unweathered material, or to expose sub-surface compositional heterogeneity, so such heterogeneity is consistent with our model. 
Such surface heterogeneity has also been detected on other spheroidal bodies, notable examples being 1999 JU3 \cite{moskovitzJU3} and Bennu \cite{binzel_bennu}. We also note the thermal observations of Delbo et al. \cite{delbo} which have shown that binary primaries have higher thermal inertias on average, perhaps indicating a loss of mass due to these phenomenon. 

\subsection{Future Work and Open Questions}

The model developed here and its application is, in many ways, simplistic and approximate. Thus, despite the intriguing and seemingly clear results that we find, they must be pursued with greater rigor. For future work there are several specific items that must be taken to the next level of analysis. We mention them here in brief, as a motivator for additional lines of inquiry and analysis.

\paragraph{Self-Consistent Gravity}

Perhaps the major assumption used here is that the geopotential is assumed to be that of a sphere, and is not modified. Although the ultimate distortion of the failed body shape may not differ significantly from a spheroid, once the loss of material from the equator was factored in, many of these computations will be sensitive to changes in the body's total gravity field and should be accommodated. One approach to do this is outlined in \cite{minton}, and will be investigated as a direction for future improvement. Conversely, it may also be possible to construct a granular-based model (c.f.\ \cite{sanchez_ApJ}) for capturing the effect of mass redistribution on the geopotential. These questions also touch on the relation between surface slope failure and global failure of a body modeled as a cohesionless rubble pile. A self-consistent gravity field calculation can track how the changes in mass distribution affect subsequent failures, creating a closed and consistent model. 

\paragraph{Effect of Cohesion}

A significant assumption in the current analysis is that the regolith is cohesionless. It has been well documented (c.f.\ \cite{scheeres_cohesion, sanchez_MAPS}) that cohesive effects can be very important for the response and failure of regolith, and it is expected that this will change the current model in some significant ways. The most fundamental change in the analysis due to cohesion would be a modification of how the angle of repose of the surface regolith would be calculated.  Once cohesion is incorporated it is necessary to also account for the weight of regolith in determining failure conditions \cite{nedderman}. This can be incorporated analytically, but requires more complex calculations and a careful derivation of the theory for this non-standard environment. Despite this, the ultimate effect may not significantly change the final results of our analysis, but could lead to a more realistic model for how a landslide is triggered. Specifically, cohesion will allow an asteroid surface to become ``perched'' at a higher energy state, and be susceptible to a small event which could then trigger a larger-scale flow. With the cohesionless model we note that failure will occur incrementally, however the inclusion of cohesion would enable to model to account for a build-up of regolith that can be potentially mobilized and make it susceptible to a triggering event. 

\paragraph{Dynamical Flow of Regolith}

Another strong assumption in our model is that the regolith flows in a quasi-static fashion, and is not subjected to dynamic phenomenon. This can be addressed through the development of realistic simulations that capture the environment of an asteroid's surface, alluded to in \cite{sanchez_ApJ, richardson_regolith_I, murdoch_regolith_II,tancredi}, and discussed at a preliminary level in \cite{scheeresLPSC2014}. There are several different aspects that such simulations can capture, and which are not captured in our current analysis. First is the direct simulation of discrete elements interacting with each other, as opposed to reliance on more analytical continuum models for describing the granular flows. Although the continuum models capture the spatially averaged mechanics of granular flows, they cannot faithfully recreate the detailed and localized motion and inertial interactions that often play an important role in granular flows. Second, this will allow the asteroid surface environment to be more faithfully replicated. For example, our current model ignores the role of Coriolis acceleration for moving grains. However, for surface flows at high spin rates this effect should cause material to flow longitudinally, destroying the simple symmetry of our current model. Similarly, it will be able to model the degree of kinetic energy dissipation that can occur in a moving flow, allowing for better estimates of what fraction of material will be lofted above the surface of the body. 

\paragraph{Orbital Mechanics of Displaced Material}

Following lofting from the surface of the asteroid, the regolith can enter a phase where it is not condensed, and where the interactions between grains becomes less important. In this regime orbital dynamics can dominate the evolution of the grains, and will ultimately control the dynamical fate of this material. Previous work has indicated that material displaced from a binary asteroid primary's surface may actually be driven back down to the surface again, affecting angular momentum transport in the system \cite{KW4_scheeres,fahnestock_tide}. The outcome may be strikingly different in a solo-asteroid situation, or may lead to formation of a satellite \cite{walsh_nature}. The presence of cohesion can also modify the expected orbital dynamics outcome. As noted in \cite{sanchez_MAPS} the presence of cohesion may allow an asteroid to spin faster before it sheds regolith, potentially putting the regolith on a more energetic -- potentially even an escape orbit -- and thus lead to different scenarios for orbit evolution of displaced material. 

\paragraph{Self-Consistent YORP Evolution}

Finally, in the analysis of this model we invoked the idea that YORP could continue to spin up an object, even after it undergoes shedding, in order to progressively evolve the surface to more uniform slopes. At the heart of this idea is that the overall YORP coefficient of a body can remain positive across surface changes. It is important to note that the YORP effect has been found to be very sensitive to the shape of the body, first described in \cite{scheeres_itokawa_YORP} and later quantified more precisely in \cite{statler}. Whether it is reasonable for a spheroidal asteroid to maintain a positive YORP coefficient should be validated with more detailed modeling. This is a crucial issue, as once the YORP coefficient of a body becomes negative, there will be a very long time before the system will reach disruption limits again. The fact that some asteroids have a very large region of constant slopes seems to indicate an answer in the affirmative to this question, yet it must still be studied and evaluated. Also, the role of Tangential YORP \cite{golubov_TYORP} could also play a factor in this, and should be evaluated in terms of a regolith-covered surface.

\section{Conclusions}

This paper explores the basic mechanics of regolith on the surface and sub-surface of rapidly spinning spheroidal bodies. Our analysis exposes some basic features for how such bodies fail and flow, how their surfaces and shapes may be deformed, and what implications these have for the loss of surface material. The theory is limited to bodies with rotational symmetry, yet these bodies constitute a fair fraction of the known small body shape morphologies. This model generates a number of specific predictions for the shapes of spheroidal bodies that have undergone failure, and we show that some current asteroid shape models share these features. These include a preferential migration of material from the mid-latitudes towards the equatorial region, the expected loss of material at the equator as it becomes pushed above the spin-off limit for such a rapidly rotating body, and the characteristic constant slope across the mid-latitudes (also noted by Harris et al. \cite{harris_tide}) and the more spherical-like cap that may remain at the polar regions. We identify aspects of these predicted shape features for the spheroidal asteroids 1999 KW4 Alpha and 2008 EV5. An analysis of P/2013 P5 is also given, where it is estimated that similar shedding events on this asteroid may occur every $\sim$ 1000 years due to YORP spin-up.

\section*{Acknowledgements}

This research was supported by grant NNX11AP24G from NASA's PG\&G program. The author appreciates the useful and insightful reviews by the referees. 

\newpage 
\bibliographystyle{plain}
\bibliography{../../bibliographies/biblio_article,../../bibliographies/biblio_conferences,../../bibliographies/biblio_misc,../../bibliographies/biblio_books}

\end{document}